\pdfoutput=1

\documentclass[onecolumn,final]{svjour3}

\smartqed

\usepackage{graphicx,amssymb}
\usepackage{epsf,psfig}
\usepackage{txfonts}
\usepackage{natbib}

\journalname{Celestial Mechanics and Dynamical Astronomy}

\begin{document}

\title{Orbit classification in the meridional plane of a disk galaxy model with
a spherical nucleus}

\author{Euaggelos E. Zotos \and Daniel D. Carpintero}

\institute{E. E. Zotos
\at Department of Physics, School of Science, Aristotle University of Thessaloniki, \\
GR-541 24, Thessaloniki, Greece, \\
\email{evzotos@physics.auth.gr}
\and
D. D. Carpintero
\at Facultad de Ciencias Astron\'{o}micas y Geof\'{i}sicas Universidad Nacional de La Plata, \\
Paseo del Bosque S/N, 1900 La Plata, Argentina 
\email{ddc@fcaglp.unlp.edu.ar}
\and
D. D. Carpintero
\at Instituto de Astrof\'{i}sica de La Plata, Conicet-Universidad Nacional de La Plata, Paseo del Bosque S/N, \\
1900 La Plata, Argentina
}

\date{Received: 14 April 2013 / Revised: 24 May 2013 / Accepted: 7 June 2013 / Published online: 19 June 2013}

\titlerunning{Orbits in the meridional plane of a disk galaxy}

\authorrunning{E. E. Zotos \& D. D. Carpintero}

\maketitle

\begin{abstract}

We investigate the regular or chaotic nature of star orbits moving in the
meridional plane of an axially symmetric galactic model with a disk and a
spherical nucleus. We study the influence of some important parameters of the
dynamical system, such as the mass and the scale length of the nucleus, the
angular momentum or the energy, by computing in each case the percentage of
chaotic orbits, as well as the percentages of orbits of the main regular
resonant families. Some heuristic arguments to explain and justify the
numerically derived outcomes are also given. Furthermore, we present a new
method to find the threshold between chaos and regularity for both Lyapunov
Characteristic Numbers and SALI, by using them simultaneously.


\end{abstract}

\section{Introduction}
\label{intro}

Although there are loads of works about the chaoticity of orbital motions in
different galactic potentials \citep[see, e.g.][citing but a
handful]{MSAB08,MA11,Z12a,Z12b}, few of them focused in the motion on the
meridional plane of an axially symmetric potential. The study of this kind of
motion can be traced back to the works of \citet{C60} and \citet{O65,O67}. While
\citet{MM75} have studied resonant (regular) meridional plane orbits, and
\citet{M79} considered, as was common in those days, that almost any orbit in an
axially symmetric potential should obey a third isolating integral of motion
besides the angular momentum and the energy, \citet{C79} claimed that the motion
of stars in the meridional plane was one of the standing problems in galactic
dynamics where integrability and stochasticity play a role. However, few
contributions to this problem have arised so far. \citet{G87,G91} insisted in
the lines of \citet{M79}, ignoring the chaoticity, while \citet{CV86} found
chaotic motion but only when their galactic model was perturbed. \citet{GS91}
have also built orbits in the meridional plane, but, again, they focused in
regular solutions claiming that most stars move on regular orbits. In the same
line, \citet{CZD00} have also studied orbits in the meridional plane but their
scope was specifically to treat regular orbits. \citet{KC01}, on the other hand,
have studied chaotic motion in a two-dimensional logarithmic potential as
representative of the meridional plane potential of an elliptical galaxy with a
dense bulge, although it lacks the necessary centrifugal term. \citet{LS92}, in
a thorough study, analyzed the orbital content in the coordinate planes of
triaxial potentials and in the meridional plane of axially symmetric potentials,
but focusing, again, on the regular families. The chaotic motion in the
meridional plane of axially symmetric galaxies, therefore, is still an open
problem, and we will conduct here an investigation of this topic.

Knowing whether the orbits of a dynamical system are ordered or chaotic is a
first step towards the understanding of the overall behavior of the system. But
also of particular interest is the distribution of regular orbits into different
families. \cite{BS82,BS84} proposed a technique, dubbed spectral dynamics, for
this particular purpose. Later on this method has been extended and improved by
\cite{L93} and \cite{CA98}. In general terms, this method computes the Fourier
transform of the coordinates of an orbit, identifies its peaks, extracts the
corresponding frequencies, and search for the fundamental frequencies and their
possible resonances. In the present work, we shall use a similar technique in
order to classify regular orbits into different families.

The present paper is organized as follows: In Section \ref{galmod} we present
our gravitational galactic model. In Section \ref{ComMet} we describe the
computational methods we used in order to explore the nature of orbits. In the
following Section, we investigate how the basic parameters of the system
influences the character of the orbits. In Section \ref{anres} we present some
heuristic arguments, in order to support and explain the numerically obtained
outcomes of the previous Section. We conclude with Section \ref{discus}, where
the discussion and the conclusions of this research are presented.

\section{The galactic model}
\label{galmod}

In the present work, we shall investigate the character of the motion in the
meridional plane of an axially symmetric disk galaxy with a spherical nucleus.
We use cylindrical coordinates $(R, \phi, z)$, where $z$ is the axis of
symmetry.

The total potential $V(R,z)$ in our model is the sum of a disk potential $V_{\rm
d}$ and a central spherical component $V_{\rm n}$. The first one is represented
by a Miyamoto-Nagai potential \citep{MN75}
\begin{equation}
V_{\rm d}(R,z) = - \frac{G M_{\rm d}}{\sqrt{R^2 + \left(\alpha + \sqrt{h^2 +
z^2}\right)^2}}.
\end{equation}
Here $G$ is the gravitational constant, $M_{\rm d}$ is the mass of the disk,
$\alpha$ is the scale length of the disk, and $h$ corresponds to the disk's
scale height. For the description of the spherically symmetric nucleus we use
a Plummer potential \citep[e.g.,][]{BT08}
\begin{equation}
V_{\rm n}(R,z) = - \frac{G M_{\rm n}}{\sqrt{R^2 + z^2 + c_{\rm n}^2}},
\end{equation}
where $M_{\rm n}$ and $c_{\rm n}$ are the mass and the scale length of the
nucleus, respectively. This potential has been used in the past to model the
central mass component of a galaxy \citep[see, e.g.][]{HN90,HPN93}.
Here we must point out that the nucleus is not intended to represent a black
hole nor any other compact object, but a bulge; therefore, we don't include
relativistic effects.

We use a system of galactic units where the unit of length is 1 kpc, the unit of
velocity is 10 km s$^{-1}$, and $G=1$. Thus, the unit of mass results $2.325
\times 10^7 {\rm M}_\odot$, that of time is $0.9778 \times 10^8$ a,\footnote{We
adhere to the recommended IAU symbol for year, i.e., ``a" \citep{W89}.} the
unit of angular momentum (per unit mass) is 10 km kpc s$^{-1}$, and the unit of
energy (per unit mass) is 100 km$^2$s$^{-2}$. We use throughout the paper the
values $M_{\rm d} = 7000$ (corresponding to $1.63\times 10^{11}$ M$_\odot$, i.e.,
a normal disc galaxy mass), $\alpha = 3$ and $h = 0.175$. These last values
were chosen having in mind a Milky Way-type galaxy \citep[e.g.,][]{AS91}.
The mass and the scale length of the nucleus, on the other hand, are treated as
parameters.

\begin{figure}
\begin{center}
\includegraphics[width=0.55\hsize]{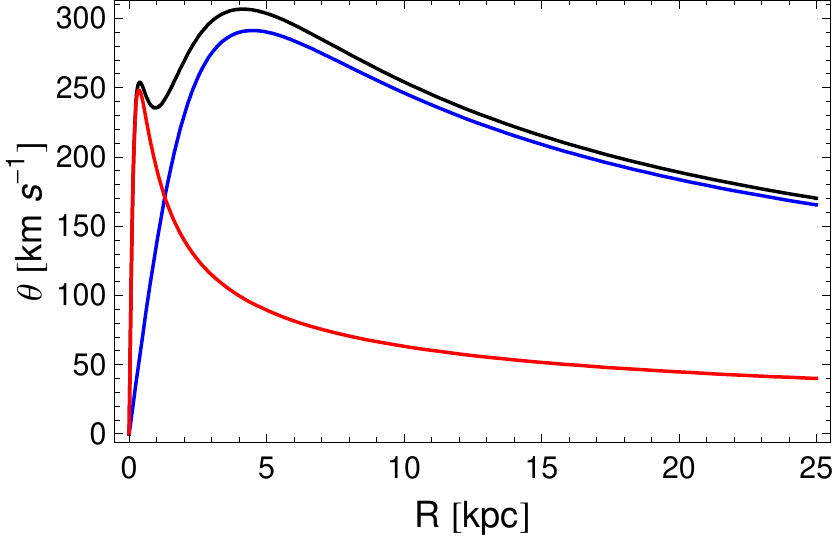}
\caption{The total circular velocity of the galactic model (black). Also shown
are the contributions from the spherical nucleus (red), and that of the disk
(blue).}
\label{rotvel}
\end{center}
\end{figure}

One important physical quantity in disk galaxies is the circular velocity in the
plane $z=0$,
\begin{equation}
\theta(R) = \sqrt{R\left|\frac{\partial V}{\partial R}\right|_{z=0}}.
\end{equation}
Replacing with our potential we obtain
\begin{equation}
\theta(R) = R \sqrt{\frac{M_{\rm d}}{\left(R^2 + (\alpha + h)^2\right)^{3/2}} +
\frac{M_{\rm n}}{\left(R^2 + c_{\rm n}^2\right)^{3/2}}}.
\end{equation}

A plot of $\theta(R)$ for our galactic model with $M_{\rm n} = 400$ and $c_{\rm
n} = 0.25$ is presented in Fig. \ref{rotvel}, as a black curve. In the same
plot, the red line shows the contribution from the spherical nucleus, while the
blue curve is the contribution from the disk. It is seen that at small distances
from the galactic center, $R\leq 1$ kpc, the contribution from the spherical
nucleus dominates, while at larger distances, $R > 1$ kpc, the disk contribution
is the dominant factor. We also observe the characteristic local minimum of the
rotation curve which appears when fitting observed data to a Galactic model
\citep[e.g.,][]{IWTS13,GHBL10}.

Since the total potential $V(R,z)$ is axisymmetric, the $z$-component of the
angular momentum $L_z$ is conserved. With this restriction, orbits can be
described by means of the effective potential
\citep[e.g.,][]{BT08}
\begin{equation}
V_{\rm eff}(R,z) = V(R,z) + \frac{L_z^2}{2R^2}.
\label{veff}
\end{equation}
The $L_z^2/(2R^2)$ term represents a centrifugal barrier; only orbits with small
$L_z$ are allowed to pass near the axis of symmetry. The 3D motion is thus
effectively reduced to a 2D motion in the meridional plane $(R,z)$, which
rotates non-uniformly around the axis of symmetry according to
\begin{equation}
\dot{\phi} = \frac{L_z}{R^2},
\end{equation}
where the dot indicates derivative with respect to time. The equations of motion
on the meridional plane are
\begin{eqnarray}
\ddot{R} &=& - \frac{\partial V_{\rm eff}}{\partial R}, \nonumber \\
\ddot{z} &=& - \frac{\partial V_{\rm eff}}{\partial z}.
\label{eqmot}
\end{eqnarray}

The equations governing the evolution of a deviation vector $\delta{\bf w}
\equiv (\delta R, \delta z, \delta \dot{R}, \delta \dot{z})$ which joins the
corresponding phase space points of two initially nearby orbits, needed for the
calculation of the standard indicators of chaos, are given by the variational
equations
\begin{eqnarray}
\dot{(\delta R)} &=& \delta \dot{R}, \nonumber \\
\dot{(\delta z)} &=& \delta \dot{z}, \nonumber \\
(\dot{\delta \dot{R}}) &=&
- \frac{\partial^2 V_{\rm eff}}{\partial R^2} \delta R
- \frac{\partial^2 V_{\rm eff}}{\partial R \partial z}\delta z,
\nonumber \\
(\dot{\delta \dot{z}}) &=&
- \frac{\partial^2 V_{\rm eff}}{\partial z \partial R} \delta R
- \frac{\partial^2 V_{\rm eff}}{\partial z^2}\delta z.
\label{vareq}
\end{eqnarray}

The corresponding Hamiltonian to the effective potential given in Eq.
(\ref{veff}) can be written as
\begin{equation}
H = \frac{1}{2} \left(\dot{R}^2 + \dot{z}^2 \right) + V_{\rm eff}(R,z) = E,
\label{energy}
\end{equation}
where $E$ is the numerical value of the Hamiltonian, which is conserved.
Therefore, an orbit is restricted to the area in the meridional plane satisfying
$E \geq V_{\rm eff}$.

\section{Computational methods}
\label{ComMet}

In order to study the chaoticity of our models, we chose, for each set of values
of the parameters of the potential, a grid of initial conditions in the $(R,\dot
R)$ plane, regularly distributed in the area allowed by the value of the energy.
The points of the grid were separated 0.1 units in $R$ and 0.5 units in $\dot
R$. For each initial condition, we integrated the equations of motion
(\ref{eqmot}) with a double precision Bulirsch-Stoer algorithm
\citep[e.g.,][]{PTVF92}. Each orbit was integrated for $10^4$ time units, which
corresponds to a time span of the order of hundreds of orbital periods. In all
cases, the energy integral (Eq. (\ref{energy})) was conserved better than one
part in $10^{-10}$, although for most orbits it was better than one part in
$10^{-11}$.

For our study, we want to know whether each orbit is regular or chaotic. Several
indicators of chaos are available in the literature; we chose two of them,
namely the standard Lyapunov exponents \citep[e.g.][]{J91} and the SALI
indicator \citep{SABV04}. Of course, being the Lyapunov exponents defined for
$t\to\infty$, only a finite time version of them (Lyapunov characteristic
numbers) are numerically achievable. To compute these indicators, we integrated,
along with each orbit, its corresponding variational equations (\ref{vareq})
from unitary displacements in each of the Cartesian directions of the phase
space $(R,z,\dot R,\dot z)$ of the meridional plane, thus allowing us to
compute, along with the SALI, the full set of Lyapunov exponents using a
Gram-Schmidt orthogonalization and a renormalization of the displacement vectors
at each step, following the recipe of \citet{BGGS80}.

To classify an orbit as regular or chaotic by using either the maximal Lyapunov
characteristic number (MLCN) or the SALI, a threshold value should be
established separating both types of orbit. This is a delicate issue, as these
thresholds are generally obtained by some statistical procedure or, even, by eye
inspection of plots of the indicators versus time. Besides, whereas the results
of different chaos indicators agree in general, there are also ``sticky" orbits
(i.e., chaotic orbits that behave as regular ones during long periods of time),
that may be misclassified by one or another method depending of the threshold
value used.

\begin{figure}
\begin{center}
\includegraphics[width=0.7\hsize]{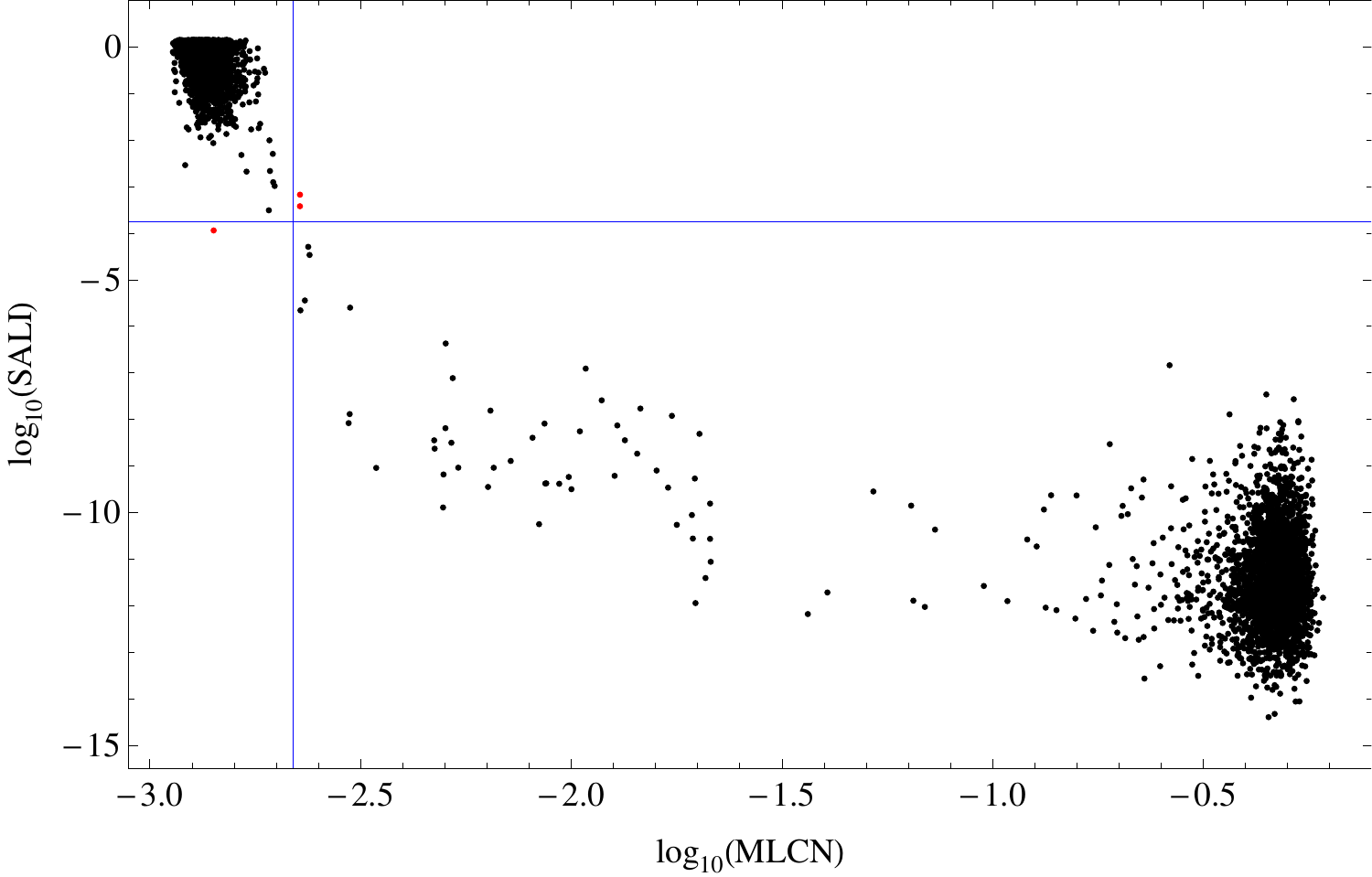}
\caption{MLCN and SALI of a set of orbits computed with $M_{\rm d} = 7000$,
$\alpha = 3$, $h = 0.175$, $c_{\rm n} = 0.25$, $M_{\rm n} =500$, $E=-670$ and
$L_z=10$. The blue lines show the resulting thresholds which separate regular
from chaotic orbits. The three red dots  indicate three orbits classified
differently by both indicators.}
\label{lysa}
\end{center}
\end{figure}

We established the thresholds by taking advantage of our computation of two
chaos indicators, as follows. First, the set of orbits of a given grid was
integrated, as we already said, for $10^4$ time units (i.e., about $10^{12}$
years, thus avoiding sticky orbits with a stickiness at least of the order of a
Hubble time). Then, their MLCN and SALI were computed, and we looked for those
values of the thresholds that maximised the agreement in the classification of
both methods. To this end, we iterated the values of the thresholds until a
minimum of disagreement has been achieved, choosing as initial values the mean
of each indicator. We found that the values thus computed leave less than 1\% of
orbits differently classified by both methods. Fig. \ref{lysa} shows an example
of the resulting thresholds. The horizontal and vertical lines correspond to the
established threshold values. We can see three red dots, corresponding to orbits
for which the classifications of both indicators did not coincide.  It is worth
noticing that \citet{KV05} also used a combination of MLCN and SALI, although
they supplemented those indicators by adding the computation of the variation
with time of the fundamental frequencies of the orbits, and they did not combine
the indicators to establish their thresholds. As a reference, Table \ref{tabla2}
shows the values of the thresholds thus obtained for each of the models studied
in Section \ref{results}, plus the number of orbits which didn't get the same
classification.

\begin{table}
\begin{center}
   \caption{Thresholds $T_{\rm S}$ and $T_{\rm L}$ obtained for the SALI and the
MLCN, respectively, total number of orbits $N$ used to compute those values, and
number $N'$ of those orbits classified differently by both indicators.}
   \label{tabla2}
   \setlength{\tabcolsep}{4.0pt}
   \begin{tabular}{@{}rrrrrrrr}
      \hline
      $M_{\rm n}$ & $c_{\rm n}$ & $L_z$ & $E$  & $\log T_{\rm S}$ &
      $\log T_{\rm L}$ & $N$ & $N'$\\
      \hline
      0-500 & 0.25      & 10   &$ -670$&$-3.80$&$-2.660$& 74050 & 11 \\
      100   & 0.05-0.50 & 10   &$ -670$&$-3.61$&$-2.689$& 64092 & 11 \\
      400   & 0.25      & 1-50 &$ -670$&$-3.46$&$-2.687$& 71425 & 13 \\
       50   & 0.25      & 10   &$ -459$&$-4.33$&$-2.653$&  8993 &  1 \\
       50   & 0.25      & 10   &$ -508$&$-3.73$&$-2.696$&  8223 &  1 \\
       50   & 0.25      & 10   &$ -567$&$-4.61$&$-2.608$&  7429 &  2 \\
       50   & 0.25      & 10   &$ -642$&$-3.00$&$-2.732$&  6575 &  2 \\
       50   & 0.25      & 10   &$ -738$&$-3.05$&$-2.701$&  5665 &  2 \\
       50   & 0.25      & 10   &$ -865$&$-2.70$&$-2.714$&  4685 &  2 \\
       50   & 0.25      & 10   &$-1038$&$-3.14$&$-2.712$&  3613 &  0 \\
       50   & 0.25      & 10   &$-1279$&$-2.75$&$-2.734$&  2473 &  0 \\
       50   & 0.25      & 10   &$-1613$&$-2.43$&$-2.736$&  1309 &  0 \\
       50   & 0.25      & 10   &$-2004$&$-2.09$&$-2.800$&   255 &  0 \\
      100   & 0.15      & 10   &$ -462$&$-2.66$&$-2.729$&  9079 &  0 \\
      100   & 0.15      & 10   &$ -511$&$-3.23$&$-2.677$&  8314 &  0 \\
      100   & 0.15      & 10   &$ -571$&$-2.96$&$-2.735$&  7506 &  0 \\
      100   & 0.15      & 10   &$ -647$&$-3.23$&$-2.719$&  6644 &  0 \\
      100   & 0.15      & 10   &$ -743$&$-2.50$&$-2.686$&  5742 &  0 \\
      100   & 0.15      & 10   &$ -871$&$-2.75$&$-2.736$&  4758 &  0 \\
      100   & 0.15      & 10   &$-1046$&$-2.90$&$-2.687$&  3703 &  0 \\
      100   & 0.15      & 10   &$-1290$&$-2.43$&$-2.729$&  2550 &  0 \\
      100   & 0.15      & 10   &$-1630$&$-2.22$&$-2.693$&  1362 &  0 \\
      100   & 0.15      & 10   &$-2037$&$-2.12$&$-2.789$&   320 &  0 \\
      500   & 0.25      & 10   &$ -489$&$-3.99$&$-2.674$&  9592 &  1 \\
      500   & 0.25      & 10   &$ -541$&$-2.65$&$-2.726$&  8812 &  2 \\
      500   & 0.25      & 10   &$ -605$&$-3.99$&$-2.659$&  7991 &  2 \\
      500   & 0.25      & 10   &$ -685$&$-3.31$&$-2.726$&  7111 &  0 \\
      500   & 0.25      & 10   &$ -788$&$-3.57$&$-2.741$&  6189 &  0 \\
      500   & 0.25      & 10   &$ -925$&$-2.75$&$-2.734$&  5188 &  0 \\
      500   & 0.25      & 10   &$-1113$&$-2.42$&$-2.751$&  4111 &  0 \\
      500   & 0.25      & 10   &$-1379$&$-2.94$&$-2.753$&  2960 &  0 \\
      500   & 0.25      & 10   &$-1763$&$-2.74$&$-2.733$&  1755 &  0 \\
      500   & 0.25      & 10   &$-2300$&$-2.00$&$-2.744$&   642 &  0 \\
      \hline
   \end{tabular}
\end{center}
\end{table}

Once the orbits have been classified into chaotic or non-chaotic, we considered
those of this last set as regular orbits.\footnote{A regular orbit of a
$N$-dimensional potential obeys, by definition, $N$ or more isolating integrals
of motion. On the other hand, a chaotic orbit is defined through its sensitivity
to the initial conditions in phase space: if the initial conditions of the orbit
are infinitesimally displaced, then the distance between the original orbit and
the new orbit grows exponentially with time. These definitions do not complement
each other. Whereas it can be proved that a regular orbit is not chaotic and a
chaotic orbit is not regular \citep[e.g.,][Section 8.3]{J91}, as far as we know
it has not been proved that every irregular (i.e. not regular) orbit is chaotic,
or, in other words, that every orbit obeying less than $N$ isolating integrals
has sensitivity to the initial conditions. Nevertheless, to avoid confusion, we
will follow here the widespread convention of considering irregular orbits and
chaotic orbits as the same set.} We then further classified these regular orbits
into families, by using the frequency analysis of \citet{CA98}, although the
extraction of frequencies was done with the frequency modified Fourier
transform, an algorithm first developed by \citet{L88} and later improved by
\citet{SN96}.

A note about the nomenclature of orbits. All the orbits of an axisymmetric
potential are 3D loop orbits, i.e., orbits that rotate around the axis of
symmetry always in the same direction. However, in dealing with the meridional
plane the rotational motion is lost, so the path that the orbit follows onto
this plane can take any shape, depending on the nature of the orbit. We will
call an orbit according to its behaviour in the meridional plane. Thus, if for
example an orbit is a rosette lying in the equatorial plane of the axisymmetric
potential, it will be a linear orbit in the meridional plane, etc.

\section{Results}
\label{results}

In this Section, we will complement our indicators of chaos, MLCN and SALI,
with the classical method of the $(R,\dot{R}, z = 0,\dot{z} > 0)$ Poincar\'e
Surface of Section (PSS), in order to visually distinguish the regular or
chaotic nature of motion. We used the initial conditions mentioned in Sec.
\ref{ComMet} in order to build the respective PSSs, taking values inside the
Zero Velocity Curve (ZVC) defined by
\begin{equation}
\frac{1}{2} \dot{R}^2 + V_{\rm eff}(R,0) = E.
\label{zvc}
\end{equation}

\begin{figure*}
\resizebox{\hsize}{!}{\includegraphics{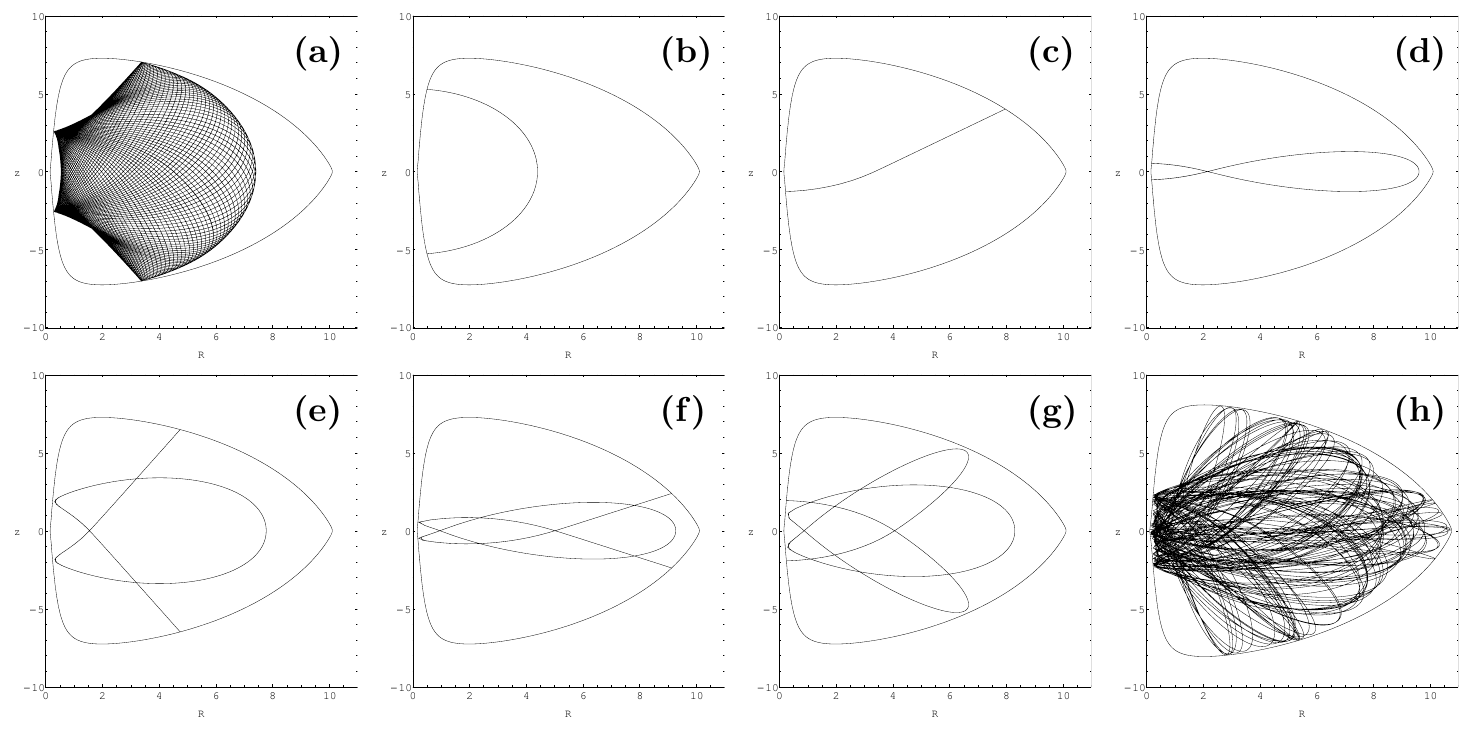}}
\caption{The eight basic types of orbits in our galactic model:
(a) box orbit; (b) 2:1 banana orbit; (c) 1:1 linear
orbit; (d) 2:3 fish orbit; (e) 4:3 boxlet orbit; (f) 4:5 boxlet orbit; (g) 6:5
boxlet orbit, one of our ``orbits with higher resonance"; (h) chaotic orbit.}
\label{orbits}
\end{figure*}

Since we want to investigate how the parameters of the dynamical system
influence not only the level of chaos but also the percentages of the basic
families of regular orbits, we chose to study the following seven basic
families: (a) box orbits, (b) 1:1 linear orbits, (c) 2:1 banana-type orbits, (d)
2:3 fish-type orbits, (e) 4:3 resonant orbits, (f) 4:5 resonant orbits and (g)
orbits with higher resonances, i.e., all resonant orbits not included in the
former categories. It turns out that for these last orbits the corresponding
percentage is less than 1\% in all cases, and therefore their contribution to
the overall orbital structure of the galaxy is insignificant. Note that every
resonance $n:m$ is expressed in such a way that $m$ is equal to the total number
of islands of invariant curves produced in the $(R,\dot{R})$ phase plane by the
corresponding orbit. In Fig. \ref{orbits} we present an example of each of the
seven basic types of regular orbits, plus an example of a chaotic one. In all
cases, we set $M_{\rm n} = 100$ (except for the chaotic orbit, where $M_{\rm n}
= 500$), $E=-670$, $L_z=10$, $c_{\rm n}=0.25$. The orbits shown in Figs.
\ref{orbits}a and \ref{orbits}h were computed until $t=100$ time units, while
the rest were computed until one period has completed. The curve circumscribing
each orbit is the limiting curve in the $(R,z)$ plane defined as $V_{\rm
eff}(R,z) = E$. Table \ref{tabla1} shows the initial conditions for each of the
depicted orbits; for the resonant cases, the initial conditions and the period
$T_{\rm per}$ correspond to the parent periodic orbit.

\begin{table}
\begin{center}
   \caption{Initial conditions for the orbits of Fig. \ref{orbits}. In all
            cases, $z_0=0$ and $\dot z_0$ is found from the energy integral,
            Eq. (\ref{energy}). $T_{\rm per}$ is the period of the orbit.}
   \label{tabla1}
   \setlength{\tabcolsep}{4.0pt}
   \begin{tabular}{@{}lcrrr}
      \hline
      Orbit      & Figure & $R_0$ & $\dot R_0$ & $T_{\rm per}$     \\
      \hline
      box        & \ref{orbits}a &  7.41800000 & 0.000000000 &          - \\
      2:1 banana & \ref{orbits}b &  4.40634102 & 0.000000000 & 1.22699769 \\
      1:1 linear & \ref{orbits}c &  3.36734581 & 31.97732133 & 0.95200932 \\
      2:3 fish   & \ref{orbits}d &  9.61363706 & 0.000000000 & 1.93809874 \\
      4:3 boxlet & \ref{orbits}e &  7.78008196 & 0.000000000 & 3.67094043 \\
      4:5 boxlet & \ref{orbits}f &  9.27613711 & 0.000000000 & 3.85375364 \\
      6:5 boxlet & \ref{orbits}g &  8.30441195 & 0.000000000 & 5.60252420 \\
      chaotic    & \ref{orbits}h & 10.72000000 & 0.000000000 &          - \\
      \hline
   \end{tabular}
\end{center}
\end{table}

It is worth noticing that the 1:1 resonance is usually the hallmark of loop
orbits, both coordinates oscillating with the same frequency in their main
motion. Their mother orbit is a closed loop orbit. Moreover, when the
oscillations are in phase, the 1:1 orbit degenerates into a linear orbit (the
same as in Lissajous figures made with two oscillators). In our meridional
plane, however, 1:1 orbits do not have the shape of a loop. Their mother orbit
is linear (as in Fig. \ref{orbits}c), and thus they don't have a hollow (in the
meridional plane) but fill a region around the linear mother, always oscillating
in $R$ and $z$ with the same frequency. We will call them ``1:1 linear open
orbits" to differentiate them from true meridional plane loop orbits, which have
a hollow and rotate always in the same sense.

\subsection{Influence of the mass of the nucleus}

To study how the mass of the nucleus $M_{\rm n}$ influences the level of chaos,
we let it vary while fixing all the other parameters of our model. As already
said, we fixed the values $M_{\rm d} = 7000$, $\alpha = 3$ and $h = 0.175$. We
chose $c_{\rm n} = 0.25$ as a fiducial value for the size of the nucleus, and
integrate orbits in the meridional plane for the set $M_{\rm n} =
\{0,50,100,...,500\}$. In all cases the energy was set to $-670$ and the angular
momentum of the orbits $L_z=10$. Once the values of the parameters were chosen,
we computed a set of initial conditions as described in Sec. \ref{ComMet}, and
integrated the corresponding orbits computing at the same time the two chaos
indicators. The respective thresholds and resulting classifications were
computed once the entire set was fully integrated.

\begin{figure*}
\resizebox{\hsize}{!}{\includegraphics{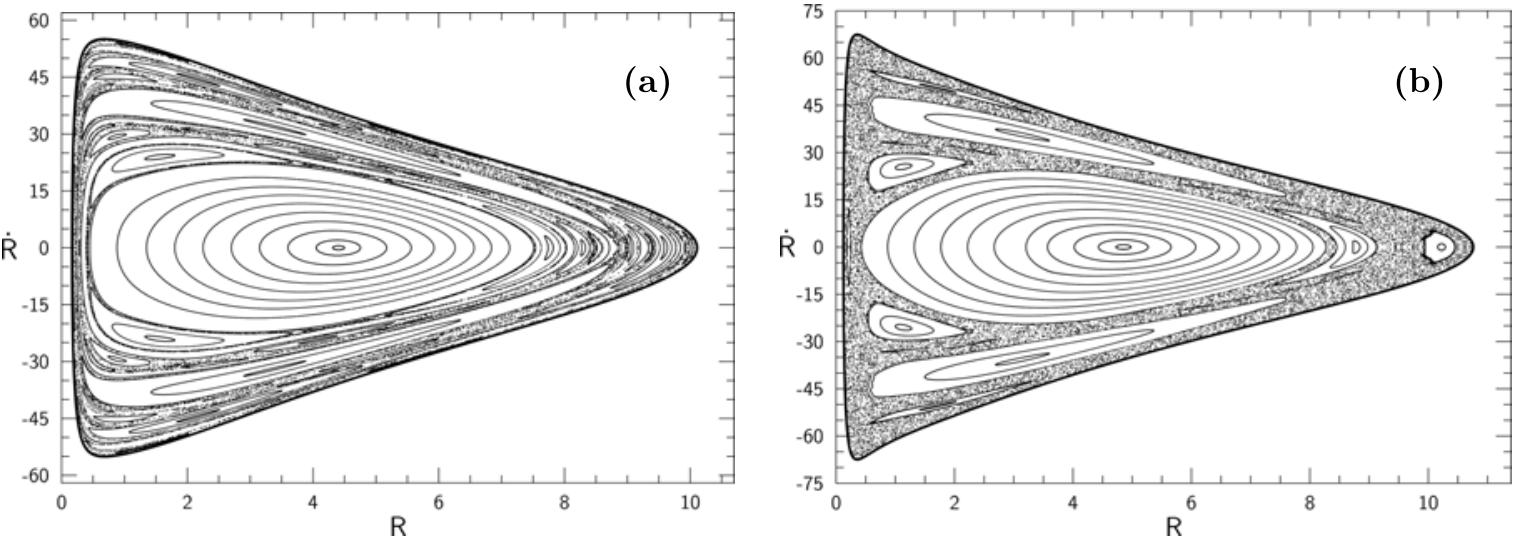}}
\caption{The $(R,\dot{R})$ phase plane when (a) $M_{\rm n} = 100$
         and (b) $M_{\rm n} = 500$.}
\label{PSSsMn}
\end{figure*}

Fig. \ref{PSSsMn}a depicts the phase plane when $M_{\rm n} = 100$. One can
observe that most of the phase space is covered by regular orbits, while there
are also several chaotic layers which separate the areas of regularity. Thus,
there is not a unified chaotic domain, at least in this $z=0$ slice of the phase
space. The outermost thick curve is the ZVC. In Fig. \ref{PSSsMn}b we present
the phase plane when $M_{\rm n} = 500$, i.e., a model with a more massive
nucleus. It is evident that there are many differences with respect to Fig.
\ref{PSSsMn}a, being the most visible the growth of the region occupied by
chaotic orbits, the  presence of a large unified chaotic sea, and an increasing
in the allowed radial velocity $\dot{R}$ of the stars near the center of the
galaxy.

\begin{figure*}
\resizebox{\hsize}{!}{\includegraphics{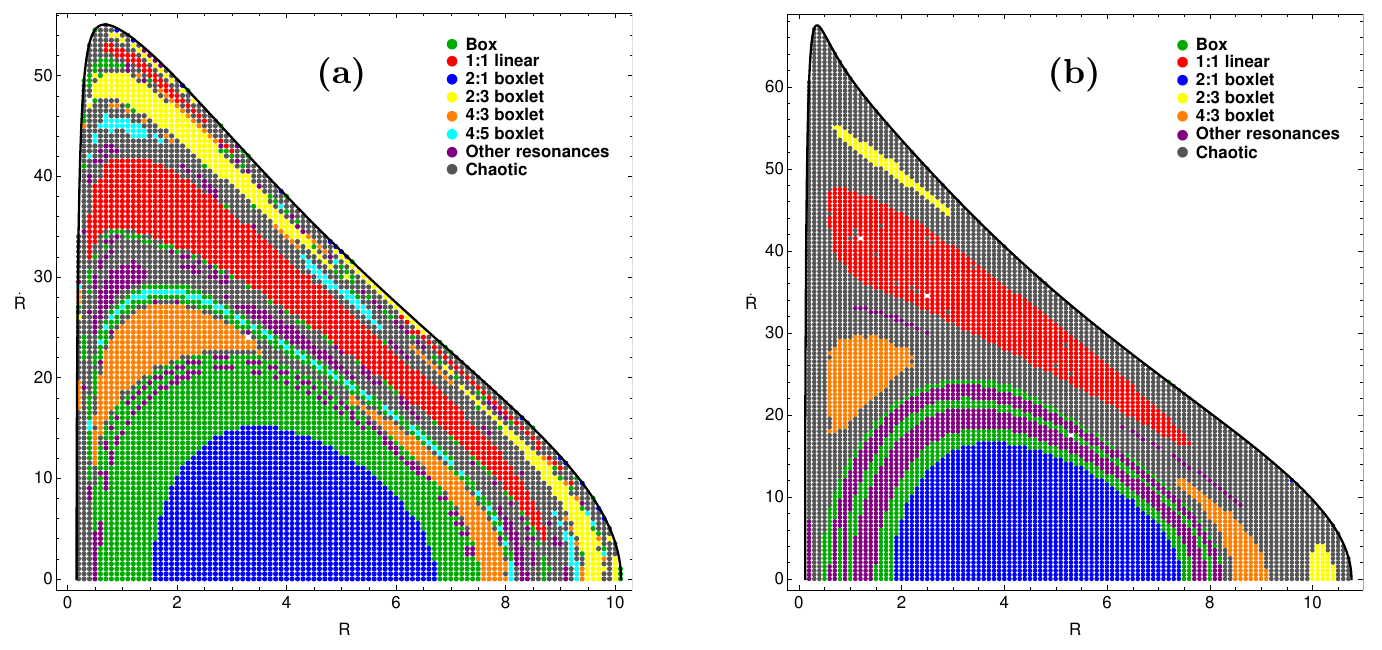}}
\caption{Orbital structure of the $(R,\dot{R})$ phase plane when
         (a) $M_{\rm n} = 100$ and (b) $M_{\rm n} = 500$.}
\label{gridMn}
\end{figure*}

Figs. \ref{gridMn}a and \ref{gridMn}b show grids of orbits that we have
classified on the PSS of Figs. \ref{PSSsMn}a and \ref{PSSsMn}b, respectively.
Here we can see which of the regular families each of the islands seen in the
PSSs belong to. In Fig. \ref{gridMn}a appear the seven main families already
mentioned: (i) 2:1 banana-type orbits correspond to the invariant curves
surrounding the central periodic point in the corresponding PSS; (ii) box orbits
are situated mainly outside of the 2:1 resonant orbits; (iii) 1:1 open linear
orbits form the double set of elongated islands in the PSS; (iv) 2:3 fish-type
orbits form the outer triple set of islands of the PSS; (v) 4:3 resonant orbits
correspond to the middle triple set of islands in the PSS; (vi) 4:5 resonant
orbits form the chain of five islands of the PSS; and (vii) there are many other
types of resonances producing several chains of small islands in the PSS,
embedded in the chaotic layers. The white dots inside the grid correspond to
orbits that have been classified as regular by one of the indicators and chaotic
by the other, being most probably sticky orbits. The outermost black thick curve
is the ZVC. On the other hand, in Fig. \ref{gridMn}b the basic families of
orbits are still present except for the 4:5 family which has disappeared making
room for chaotic orbits. Also, the portion of higher resonant orbits is
considerably smaller than in Fig. \ref{gridMn}a.

\begin{figure}
\begin{center}
\includegraphics[width=0.6\hsize]{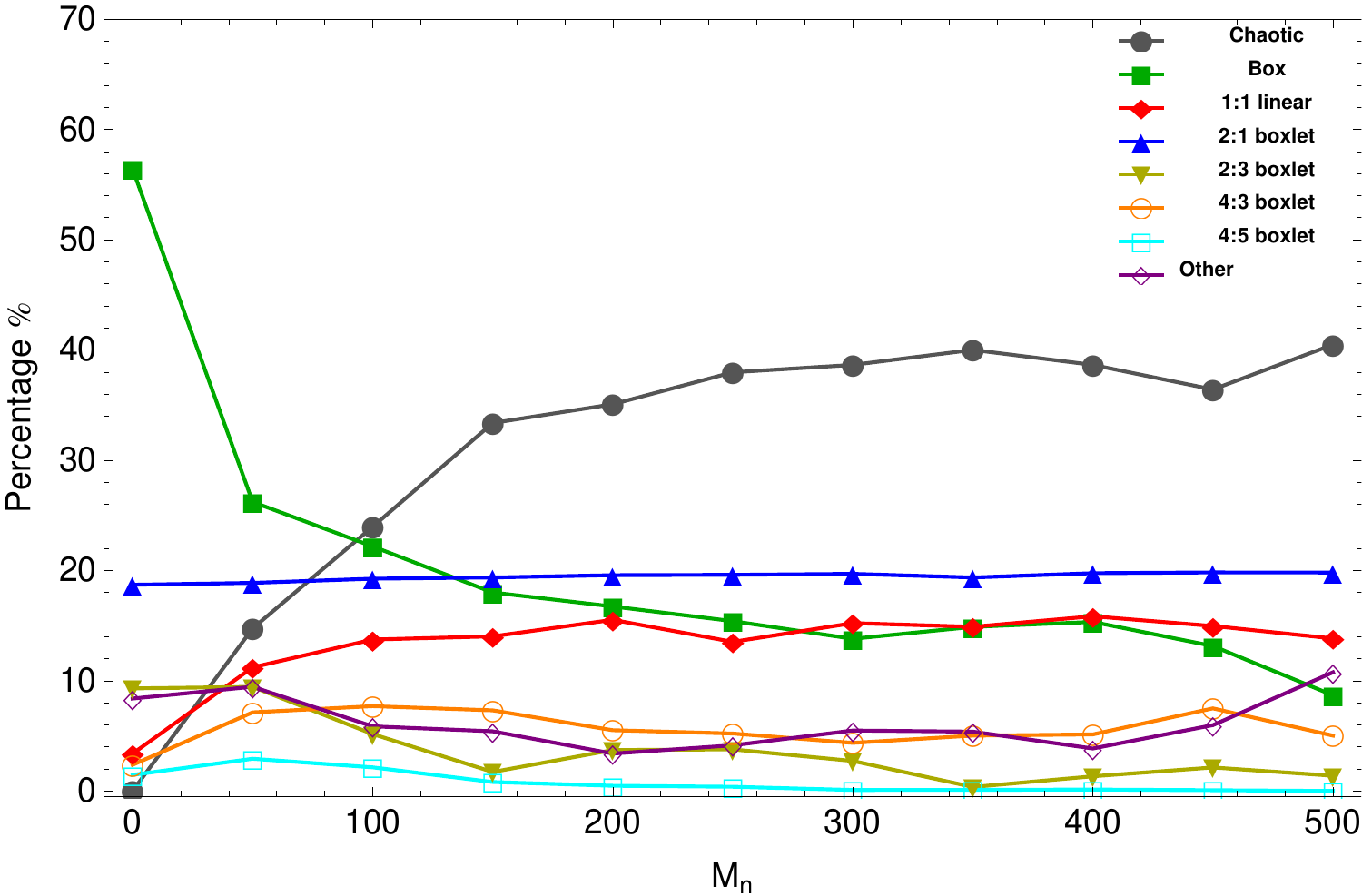}
\caption{Percentages of different kinds of orbits, varying $M_{\rm n}$.}
\label{percMn}
\end{center}
\end{figure}

Fig. \ref{percMn} shows the resulting percentages of chaotic orbits and of the
mean families of regular orbits as $M_{\rm n}$ varies. It can be seen that when
the nucleus is absent, there is no chaos whatsoever, and most orbits are box
orbits. However, a small nucleus is enough to trigger chaotic phenomena, whereas
the box orbits are depleted. This trend continues, although at a lesser rate, as
the nucleus grows in mass, i.e., the percentage of box orbits is reduced and
that of chaotic orbits is increased. The rest of orbits change very little; the
meridional bananas, in fact, are almost unperturbed by the shifting of the mass
of the nucleus. From this figure, one may conclude that $M_{\rm n}$ affects
mostly the box and chaotic orbits in our galactic model.

\subsection{Influence of the scale length of the nucleus}

Now we proceed to investigate how the scale length of the nucleus $c_{\rm n}$
influences the amount of chaos. Again, we let it vary while fixing all the other
parameters of our galactic model, choosing $M_{\rm n} = 100$ as a fiducial value
for the mass of the nucleus, and integrating orbits in the meridional plane for
the set $c_{\rm n} = \{0.05,0.10,0.15,...,0.50\}$. As before, the energy was set
to $-670$ and $L_z = 10$.

\begin{figure*}
\resizebox{\hsize}{!}{\includegraphics{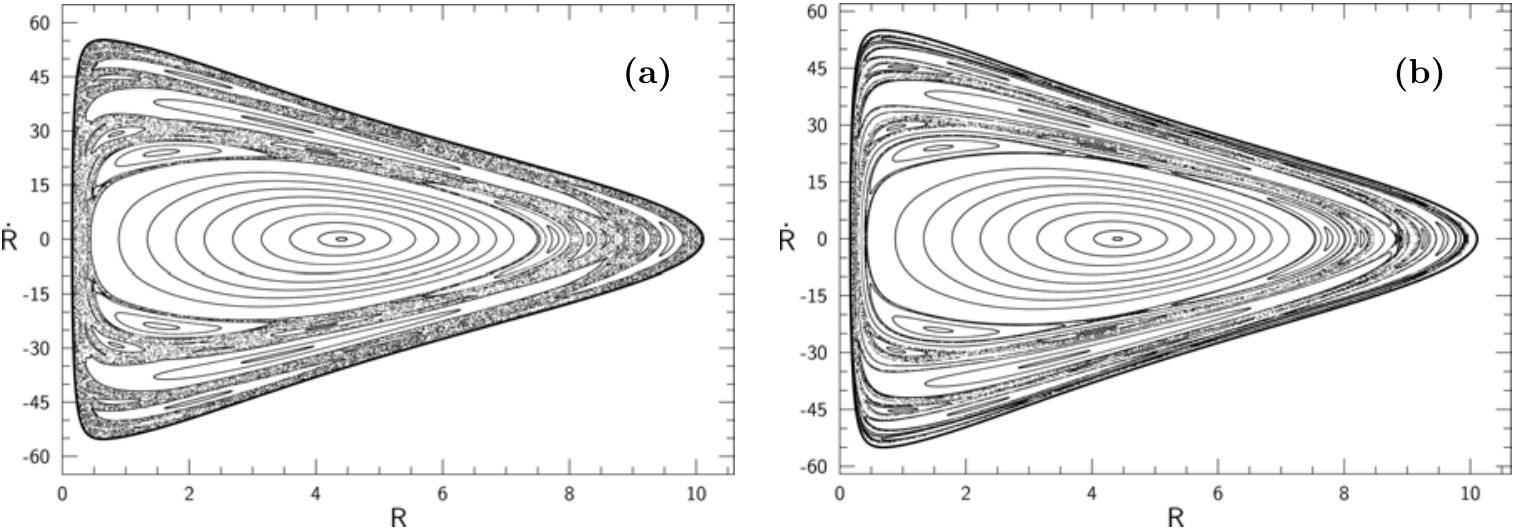}}
\caption{The $(R,\dot{R})$ phase plane when (a) $c_{\rm n} = 0.05$
         and (b) $c_{\rm n} = 0.30$.}
\label{PSSscn}
\end{figure*}

\begin{figure*}
\resizebox{\hsize}{!}{\includegraphics{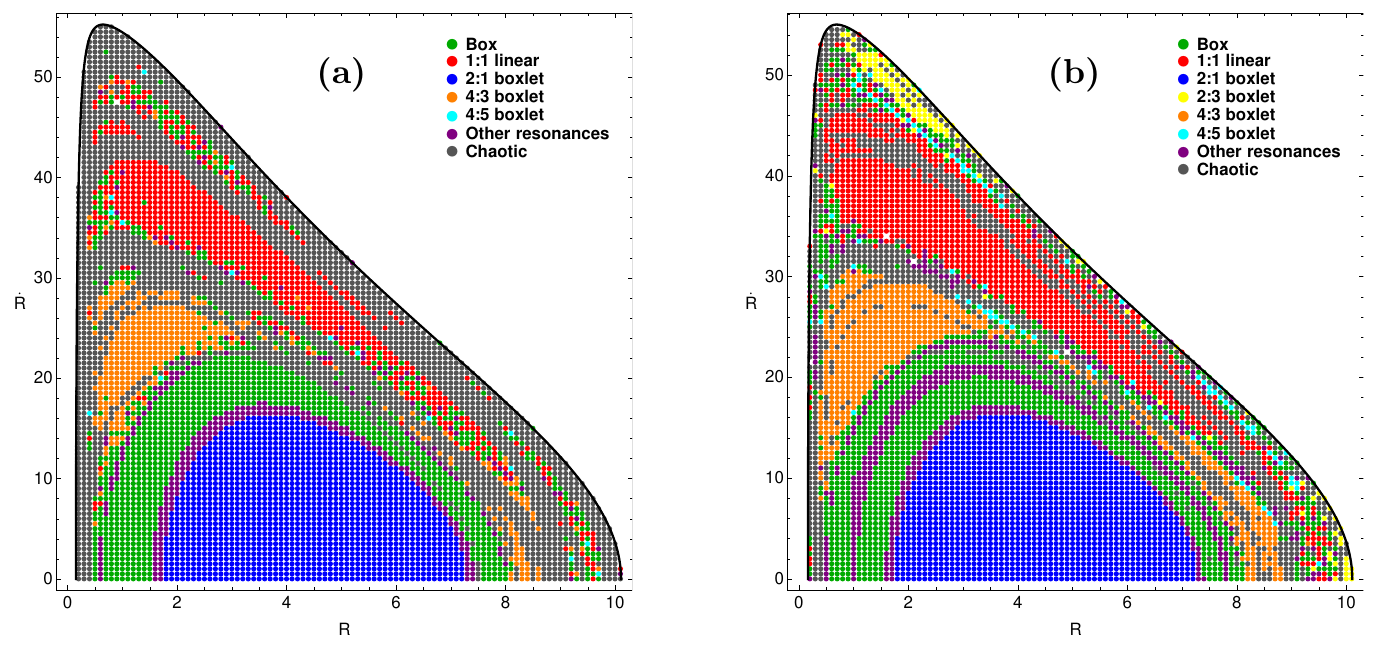}}
\caption{Orbital structure of the $(R,\dot{R})$ phase plane when
         (a) $c_{\rm n} = 0.05$ and (b) $c_{\rm n} = 0.30$.}
\label{gridcn}
\end{figure*}

\begin{figure}
\begin{center}
\includegraphics[width=0.6\hsize]{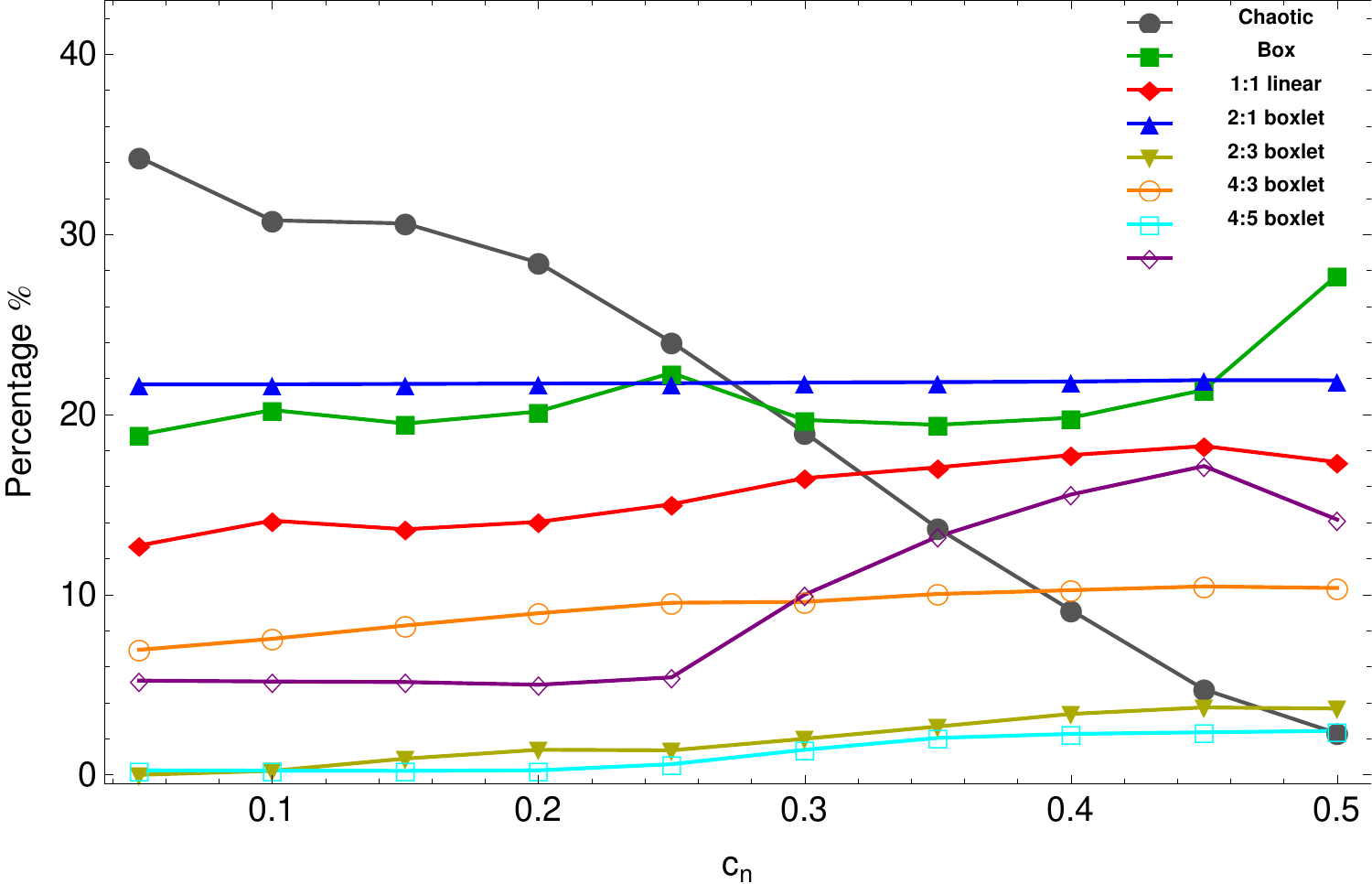}
\caption{Percentages of different kinds of orbits, varying $c_{\rm n}$.}
\label{perccn}
\end{center}
\end{figure}

Fig. \ref{PSSscn}a presents the phase plane when $c_{\rm n} = 0.05$. We observe
that much of it is covered by islands of regular orbits, surrounded by a chaotic
sea. Fig. \ref{PSSscn}b corresponds to the $c_{\rm n} = 0.30$ case, i.e., a less
concentrated spherical nucleus. As can be seen, there are no big differences
between both cases, the most prominent one being the shrinking of the area
occupied by chaotic orbits. In fact, the chaotic sea has been separated into
different regions which surround all the sets of invariant curves produced by
the regular orbits. Fig. \ref{gridcn}a, made as Fig. \ref{gridMn}a but for the
PSS of Fig. \ref{PSSscn}a, shows that there are present only six out of the
seven main families of regular orbits: the 2:3 resonance is absent, while the
islands of the 4:5 resonant orbits are so thin that they appear as lonely points
in the grid. On the other hand, in Fig. \ref{gridcn}b, corresponding to $c_{\rm
n} = 0.30$, all the basic families of regular orbits are present. Also, the
region of orbits of high resonances is considerably larger than with $c_{\rm n}
= 0.05$. This trend is fairly visible in Fig. \ref{perccn}, where the resulting
percentages of chaotic and regular orbits as $c_{\rm n}$ varies are shown. It
can be seen that there is a strong correlation between the percentage of chaotic
orbits and the value of $c_{\rm n}$. At the same time, as the nucleus become
less concentrated, there is a gradual increase in the percentage of almost all
of the regular families, most noticeably the box and the high resonant boxlets.
Once again, the meridional 2:1 bananas are immune to changes of the parameter.
Thus, decreasing the scale length of the nucleus turns box and high resonant
orbits into chaotic orbits, while those with low resonances are less affected.

\subsection{Influence of the angular momentum}

\begin{figure*}
\resizebox{\hsize}{!}{\includegraphics{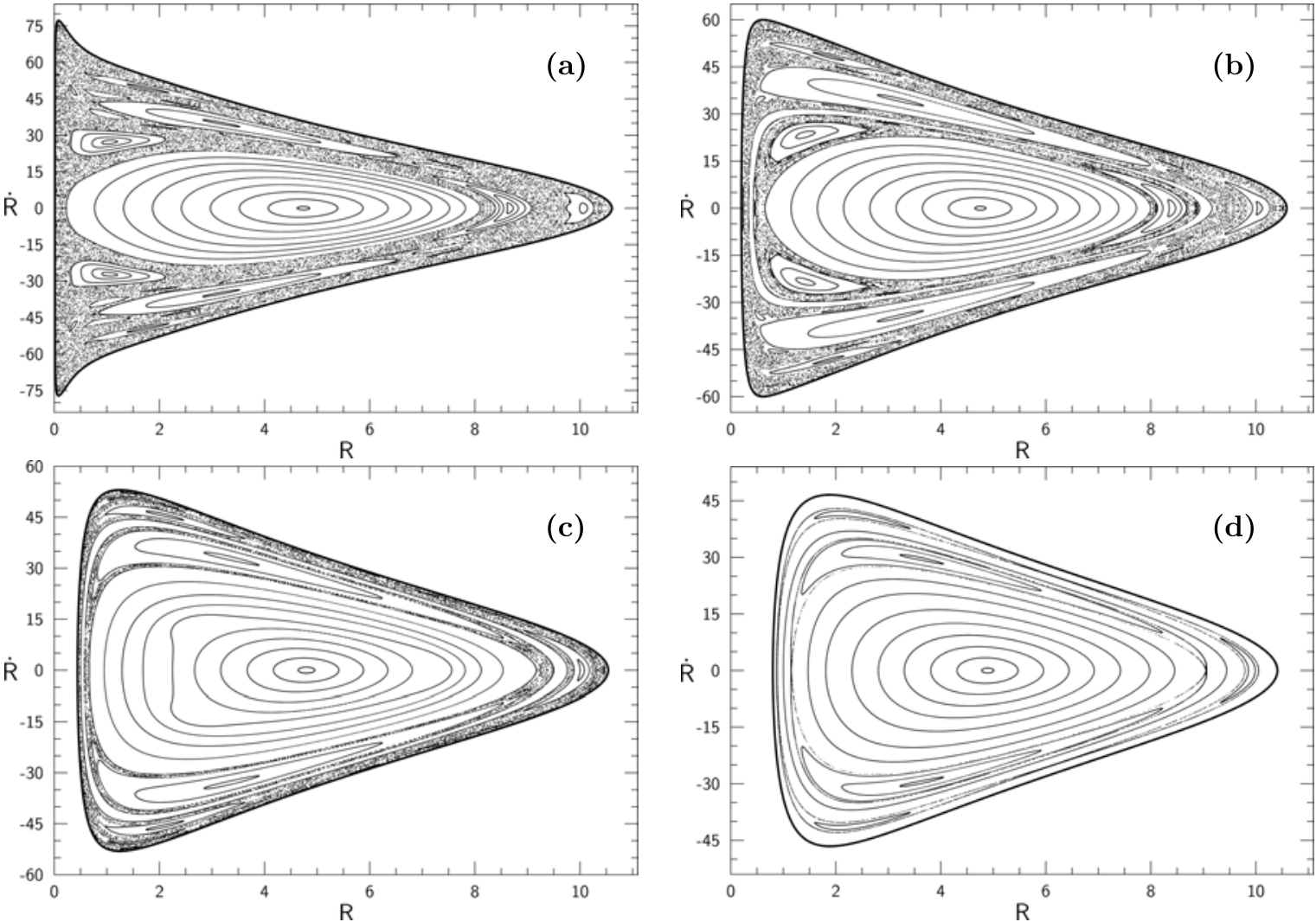}}
\caption{The $(R,\dot{R})$ phase plane when (a) $L_z = 10$,
         (b) $L_z = 15$, (c) $L_z = 30$ and (d) $L_z = 50$.}
\label{PSSsLz}
\end{figure*}

\begin{figure*}
\resizebox{\hsize}{!}{\includegraphics{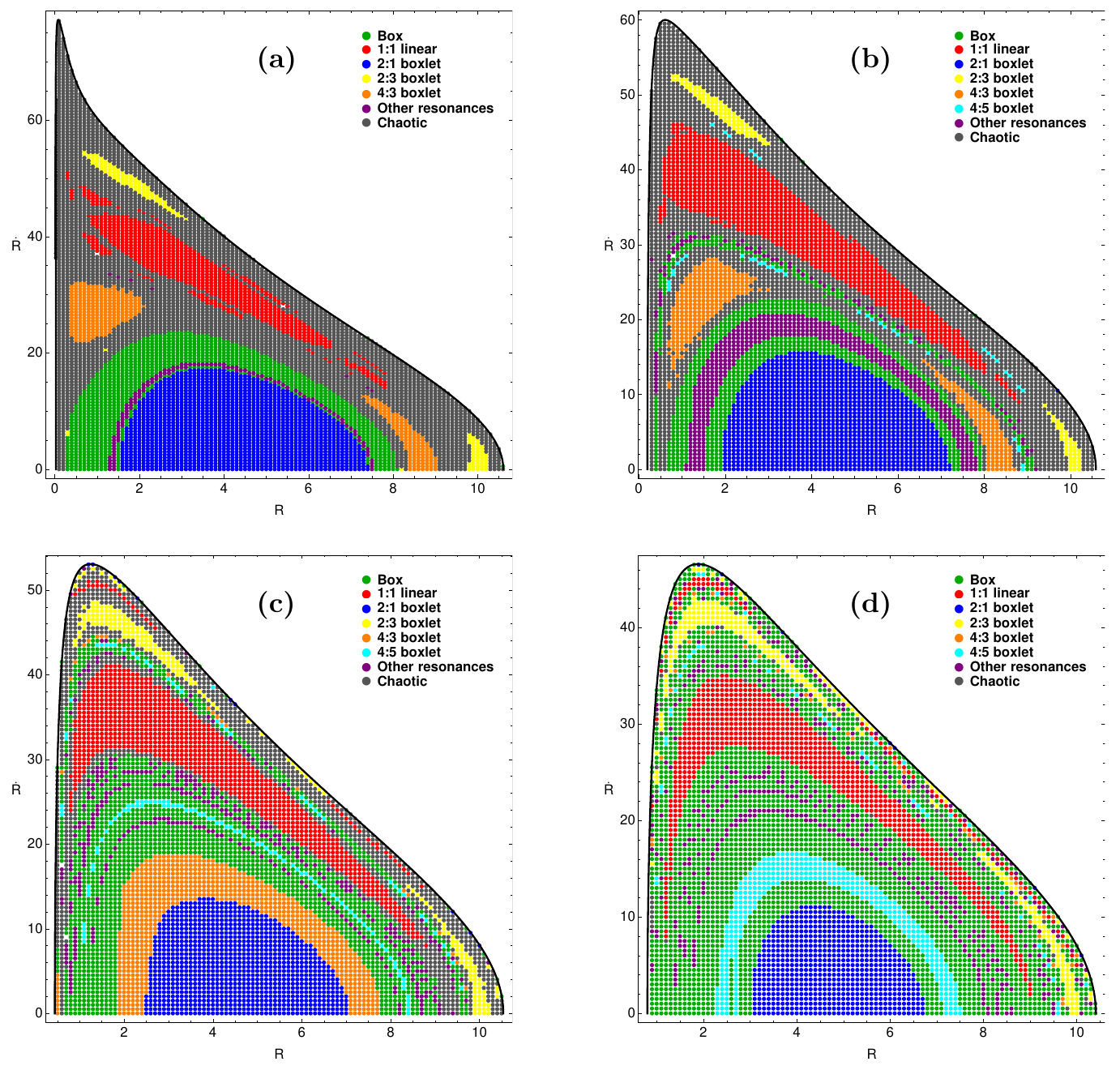}}
\caption{Orbital structure of the $(R,\dot{R})$ phase plane when (a)
  $L_z = 1$, (b) $L_z = 15$, (c) $L_z =
         30$ and (d) $L_z = 50$.}
\label{gridLz}
\end{figure*}

\citet{Z12b} showed, for an axially symmetric galactic model composed of a disk,
a halo and a spherical nucleus, that one of the most important parameters that
influences the orbital structure is the angular momentum $L_z$. Here, we let
$L_z$ vary along the set $\{1,5,10,15,...,50\}$, while fixing $M_{\rm n} = 400$,
$c_{\rm n} = 0.25$ and $E=-670$. Fig. \ref{PSSsLz}a depicts the $(R,\dot{R})$
phase plane when $L_z = 1$. We observe the existence of a large chaotic sea,
while most of the regular orbits are located near the central region of the
phase plane, although there are important islands of invariant curves
surrounding them. In Fig. \ref{PSSsLz}b, corresponding to $L_z = 15$, we can see
that the amount of chaos is smaller, and that an additional inner chaotic layer
has appeared besides the outer sea. In Fig. \ref{PSSsLz}c we can see the phase
plane when $L_z = 30$. It is seen that almost all the phase plane is covered by
regular orbits. Nevertheless, we can still distinguish the presence of two
distinct chaotic layers. In Fig. \ref{PSSsLz}d, which shows the case $L_z = 50$,
the entire phase plane is seen covered by regular orbits; the chaotic motion is
negligible. From these figures we can draw two conclusions: (i) increasing $L_z$
causes a decreasing of the chaotic region, which eventually disappears almost
completely and (ii) the permissible area on the $(R,\dot{R})$ phase plane is
reduced as we increase the value of $L_z$. Figs. \ref{gridLz}(a-d) show the
grids of orbits that we have classified, corresponding to the PSSs of Figs.
\ref{PSSsLz}(a-d) respectively. In Fig. \ref{gridLz}a we note a lack of 4:5
boxlets orbits; they appear in Fig. \ref{gridLz}b, although forming very thin
islands. In Fig. \ref{gridLz}c we observe a drastic decrease of chaotic orbits,
thus leaving room to regular families. Fig. \ref{gridLz}d shows that the regular
orbits have taken almost all the phase plane. It is worth noticing that, at the
highest angular momentum, the 4:3 resonance has been almost deleted from the
phase plane.

\begin{figure*}
\resizebox{\hsize}{!}{\includegraphics{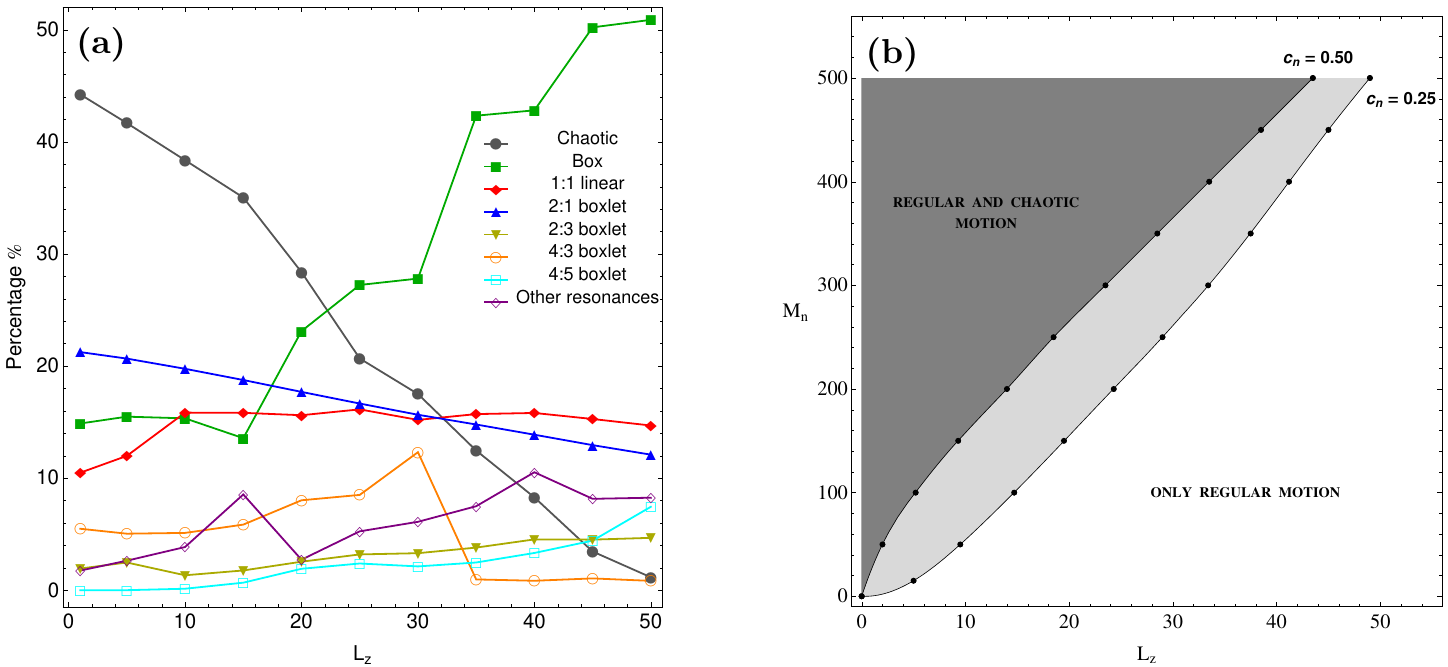}}
\caption{(a): Percentages of different kinds of orbits, varying $L_z$.
(b): Relationship between the critical value of the angular momentum
$L_{z\rm c}$ and the mass of the nucleus $M_{\rm n}$, for two values of the
scale length of the nucleus $c_{\rm n}$. More details are given in the text.}
\label{Lzpercrit}
\end{figure*}

Fig. \ref{Lzpercrit}a presents the resulting percentages of chaotic and
regular orbits as $L_z$ varies. It is clearly seen that, as $L_z$ increases, the
percentage of chaotic orbits decreases almost linearly, while that of box orbits
grows steadily when $L_z>15$. In fact, when $L_z>25$, they are the dominant type
of orbits. The rest of orbits change less. In particular, the percentage of 2:1
bananas is little affected by the increase of the angular momentum, unlike the
previous cases. It is also seen that when $L_z \simeq 30$ the percentage of the
4:3 family drops suddenly, remaining at very low values from then on.
Summarizing, the angular momentum mostly affects box and chaotic orbits.

To further investigate the influence of $L_z$, we define, for each set of values
of the parameters, the critical value of the angular momentum $L_{z\rm c}$ as
the maximum value of the angular momentum for which the orbits can display
chaotic motion. Fig. \ref{Lzpercrit}b shows the numerically found
relationship between $L_{z\rm c}$ and $M_{\rm n}$, for two values of the scale
length of the nucleus, $c_{\rm n} = 0.25$ and $c_{\rm n} = 0.50$. To obtain
this, orbits were started near $R_0 = R_{\rm min}$, with $z_0 = \dot{R_0} = 0$,
and $\dot{z_0}$ obtained from the energy integral with $E = -670$. Here, $R_{\rm
min}$ is the minimal root of the equation
\begin{equation}
 V(R,0) + \frac{L_z^2}{2R^2} = E.
 \label{minroot}
\end{equation}

The particular value of the energy was chosen so that in all cases $R_{\rm max}
\simeq 10$ kpc. It is seen in Fig. \ref{Lzpercrit}b that the relationship
between $L_{z\rm c}$ and $M_{\rm n}$ is nearly linear. This line divides the
$(L_z, M_{\rm n})$ plane in two parts; orbits on the upper shaded part of the
plot may be either regular or chaotic, while those on the lower part of the plot
can only be regular. Moreover, it is interesting to notice that the extent of
the chaotic domain is larger when the value of $c_{\rm n}$ is smaller, that is,
when the nucleus is more concentrated, which is in agreement with the results
found in Fig. \ref{perccn}.

\subsection{Influence of the energy}

Another parameter that plays an important role is the value of the orbital
energy $E$. We investigated three cases: (i) A system with a small amount of
chaos; for this case, we set $M_{\rm n} = 50$, $c_{\rm n} = 0.25$ and $L_z =
10$. (ii) A system with a medium level of chaos, for which we chose $M_{\rm n} =
100$, $c_{\rm n} = 0.15$ and $L_z = 10$. (iii) A system which presents a large
amount of chaos, with $M_{\rm n} = 500$, $c_{\rm n} = 0.25$ and $L_z = 10$. To
select the energy levels, we chose for each case the ten energies which give
$R_{\rm max} = \{1.5,3,4.5,...,15\}$, where $R_{\rm max}$ is the maximum
possible value of $R$ on the $(R,\dot R)$ phase plane.

\begin{figure*}
\resizebox{\hsize}{!}{\includegraphics{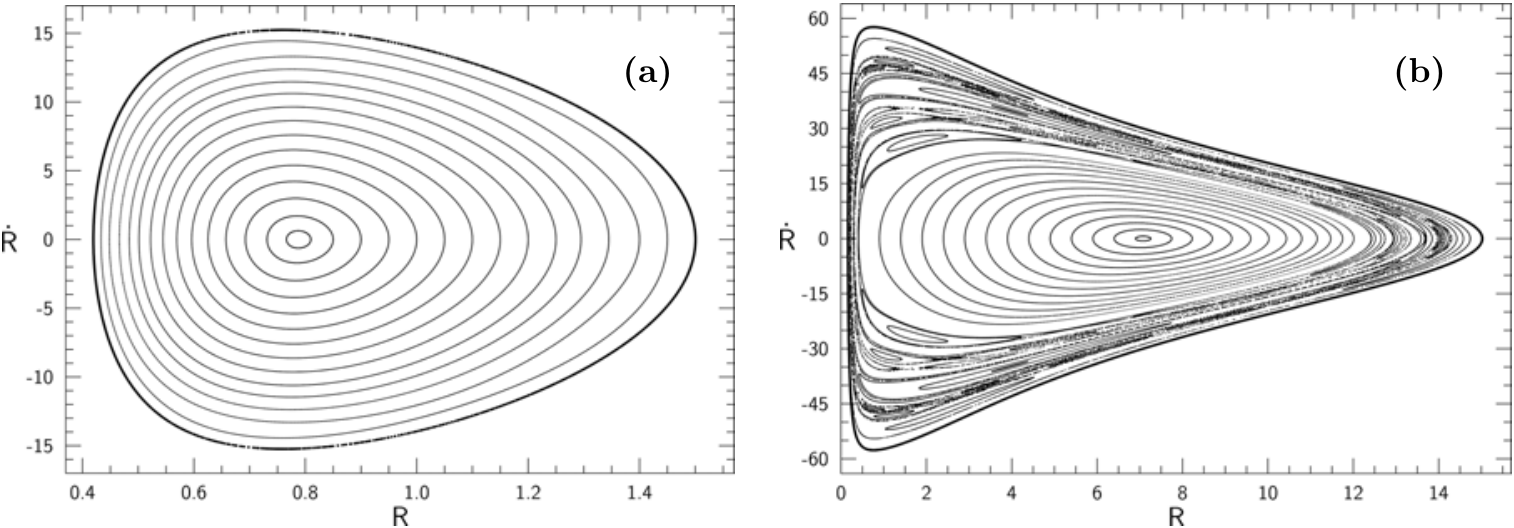}}
\caption{The $(R,\dot{R})$ phase planes of case (i) when (a) $E =
         -2004$ and (b) $E = -459$.}
\label{PSSsEnS}
\end{figure*}

\begin{figure*}
\resizebox{\hsize}{!}{\includegraphics{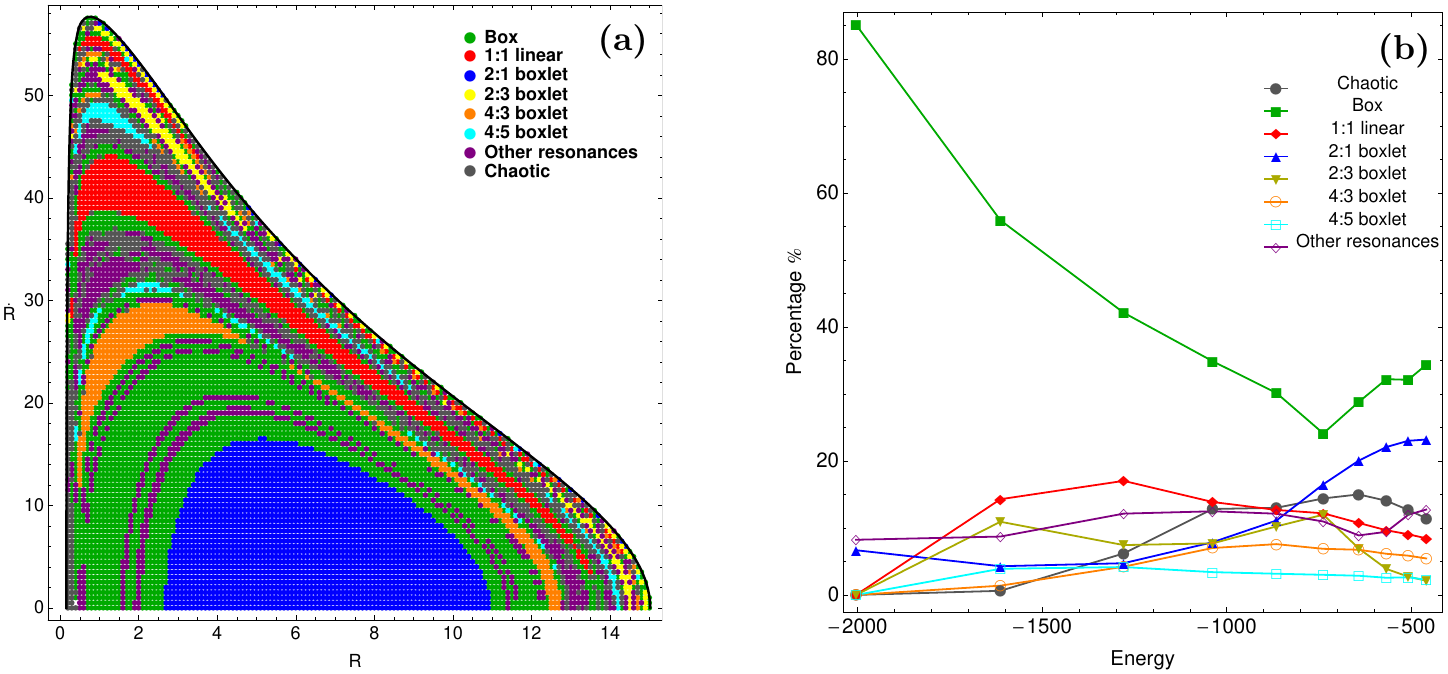}}
\caption{(a): Orbital structure of the $(R,\dot{R})$ phase plane of case (i)
when $E = -459$. (b): Percentages of different kinds of orbits, varying the
energy $E$ in case (i).}
\label{gridEnS}
\end{figure*}

For the case (i), low level of chaos, Fig. \ref{PSSsEnS}a depicts the phase
plane when $E = -2004$ which corresponds to $R_{\rm max} = 1.5$. It is seen that
the entire phase space is covered by regular orbits. In Fig. \ref{PSSsEnS}b we
present the phase plane when $E = -459$ ($R_{\rm max} = 15$). Here, the chaotic
orbits have taken a small region of the phase plane, which is nevertheless still
mainly occupied by regular families. Fig. \ref{gridEnS}a shows a grid of
orbits that we have classified on the PSS of Fig. \ref{PSSsEnS}b. We can see
that all the different types of orbits are present, although most of them
correspond to either box orbits or 2:1 banana-type orbits. Also, the 1:1 and 4:3
resonances occupy a significant portion of the PSS, while the 2:3 and the 4:5
resonances appear to be confined in small islands. Fig. \ref{gridEnS}b
shows the resulting percentages of chaotic and regular orbits as $E$ varies. It
can be seen that when the motion of stars is at very low energies, it is
entirely regular, being the box orbits the all-dominant type. The percentage of
box orbits is however reduced as the energy is increased, although they always
remain the most populated family. It is also seen that the percentages of 1:1
and 2:3 boxlets start to grow as soon as the energy grows, but then they remain
at relatively low values, the 2:3 family disappearing at high energies. The
percentage of orbits of the 2:1 family, on the other hand, takes relatively high
values at high energies. Only the 4:5 boxlet orbits remain almost unperturbed by
the increase of the energy. These percentages show that, when there is a small
amount of chaos, the value of the energy affects mostly the regular orbits by
shifting the population of the different families.

\begin{figure*}
\resizebox{\hsize}{!}{\includegraphics{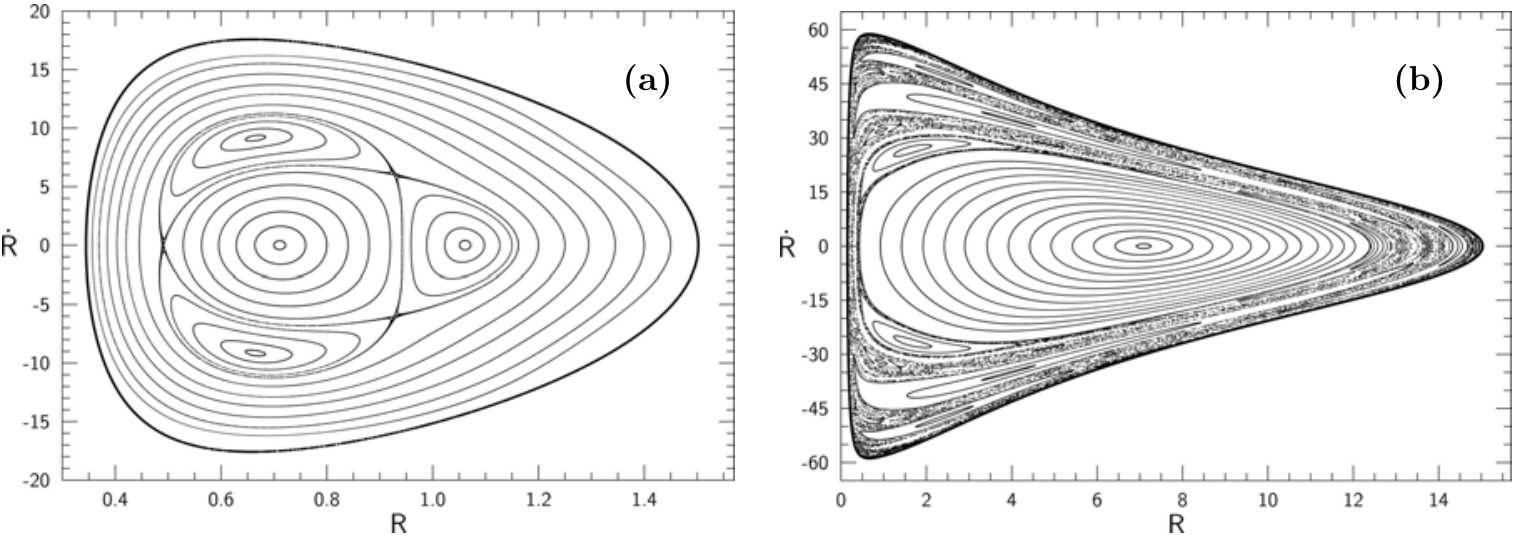}}
\caption{The $(R,\dot{R})$ phase planes of case (ii) when (a) $E =
         -2037$ and (b) $E = -462$.}
\label{PSSsEnM}
\end{figure*}

\begin{figure*}
\resizebox{\hsize}{!}{\includegraphics{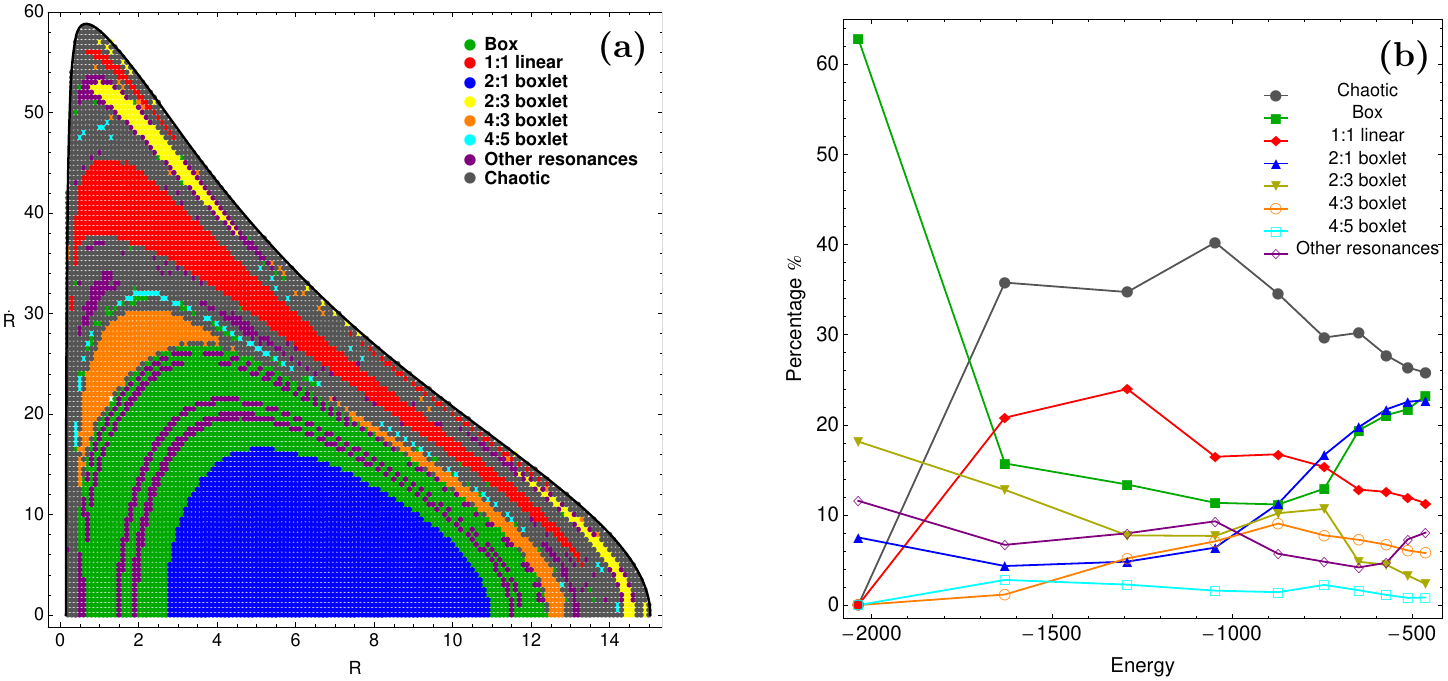}}
\caption{(a): Orbital structure of the $(R,\dot{R})$ phase plane of case (ii)
when $E = -462$. (b): Percentages of different kinds of orbits of case (ii),
varying the energy $E$.}
\label{gridEnM}
\end{figure*}

For the case (ii), Fig. \ref{PSSsEnM}a depicts the phase plane when $E = -2037$
which corresponds to $R_{\rm max} = 1.5$. Once again, the phase plane is almost
entirely covered by regular orbits, while the chaotic motion is negligible. In
Fig. \ref{PSSsEnM}b we present the phase plane when $E = -462$ and $R_{\rm max}
= 15$, where a chaotic outer area is clearly seen, and an additional inner thin
chaotic layer can be observed. Fig. \ref{gridEnM}a shows a grid of orbits
that we have classified on the PSS of Fig. \ref{PSSsEnM}b. It is seen that the
majority of orbits correspond either to chaotic, box, or 2:1 banana-type orbits.
The 1:1, 2:3 and 4:3 resonances occupy a significant portion of the PSS forming
well-defined islands. It is also worth mentioning that there is a considerable
amount of higher resonant orbits. Fig. \ref{gridEnM}b presents the
resulting percentages of the different orbits as $E$ varies. Once again, low
energy orbits, i.e., orbits which move near the center, are almost entirely box
orbits (in the meridional plane). Also, the 2:1 bananas start to grow in
percentage at high energies. But, unlike case (i), the percentage of box orbits
is significantly reduced as $E$ grows, and, at the same time, the chaotic and
the 1:1 orbits increase rapidly. The level of chaoticity remains relatively
high, around 30-40\%. At the higher energy studied in case (ii), the percentages
of the chaotic, box and 2:1 orbits tend to a common value (around 25\%), sharing
three fourths of the entire phase plane.

\begin{figure*}
\resizebox{\hsize}{!}{\includegraphics{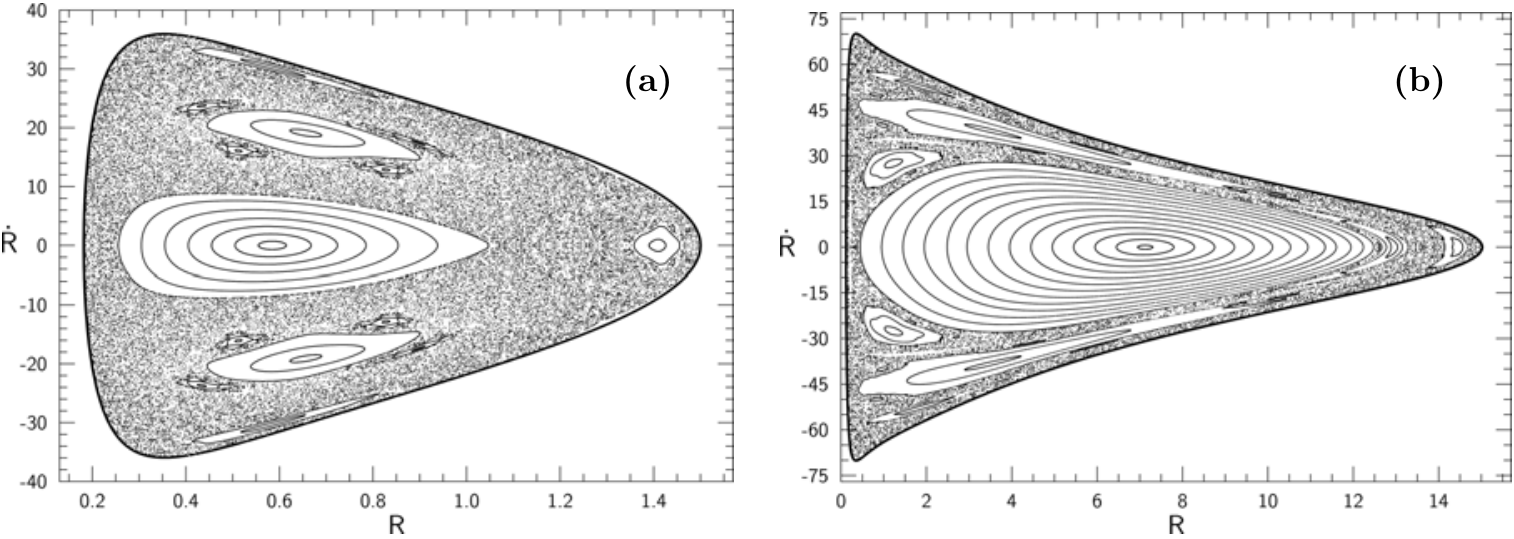}}
\caption{The $(R,\dot{R})$ phase planes of case (iii) when (a) $E =
         -2300$ and (b) $E = -489$.}
\label{PSSsEnL}
\end{figure*}

\begin{figure*}
\resizebox{\hsize}{!}{\includegraphics{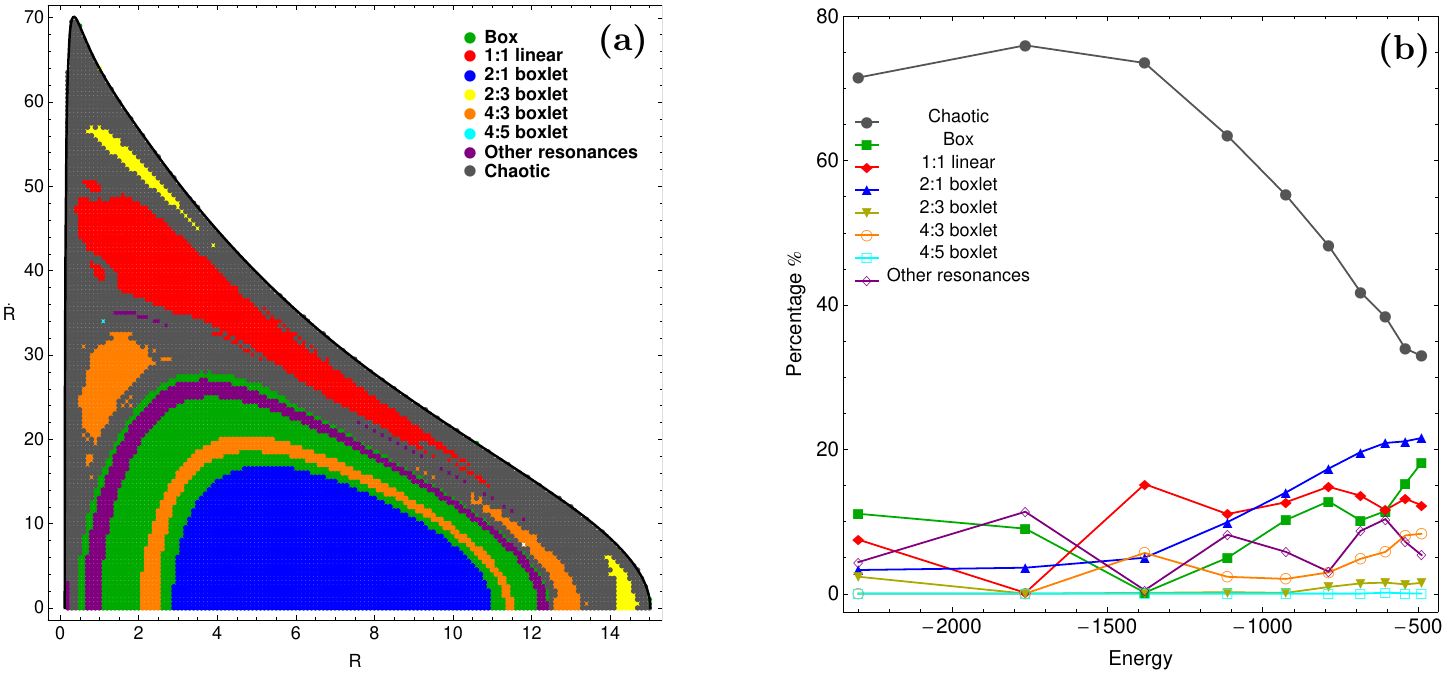}}
\caption{(a): Orbital structure of the $(R,\dot{R})$ phase plane of case (iii)
when $E = -489$. (b): Percentages of different kinds of orbits in case (iii),
varying the energy $E$.}
\label{gridEnL}
\end{figure*}

With respect to case (iii), high level of chaos, Fig. \ref{PSSsEnL}a shows the
phase plane when $E = -2300$ which corresponds to $R_{\rm max} = 1.5$. It is
seen that most of the phase plane is covered by chaotic orbits. In Fig.
\ref{PSSsEnL}b we present the phase plane when $E = -489$  and $R_{\rm max} =
15$. Here, we have a significant decrease in the region occupied by chaotic
orbits. Fig. \ref{gridEnL}a shows the grid of orbits corresponding to the
PSS of Fig. \ref{PSSsEnL}b. We see that most orbits are chaotic, although all
the regular families, with the exception of the 4:5 family, occupy a significant
portion of the PSS. In Fig. \ref{gridEnL}b, which presents the resulting
percentages, it can be seen that the motion is highly chaotic throughout. Though
the percentage of chaotic orbits is gradually reduced as the energy increases,
it remains larger than any other individual regular family. From $E \simeq
-1400$ on, the decreasing of chaotic orbits is paired with a similar increasing
of box and 2:1 orbits. As before, the 4:5 orbits remain unperturbed and with
very low percentages

\section{Analysis of the results}
\label{anres}

In the literature, the dynamical origin of the onset of chaos has proven elusive
so far. A promising line of investigation, namely the curvature of the phase
space, although theoretically sound, came up against many experimental
counterexamples \citep[e.g.,][]{S94}. So, we will not attempt to explain which
dynamical factors are responsible for the onset and growth of chaos, but try
to isolate any behaviour that may be correlated with that.

\citet{GB85} have shown that stars that pass near a density cusp, thus receiving
a large acceleration, may depopulate the family of box orbits which supports the
triaxial figure of a galaxy. Therefore, regions of large accelerations may be
responsible for the onset of chaos. Since our potential is nowhere divergent,
we do not have any cusps. But, the effective potential on the meridional plane
\emph{does} have a cusp at the origin, caused by the centrifugal term. Thus, we
seek whether there is a relationship between chaos and a star going near the
origin.

\begin{figure*}
\resizebox{\hsize}{!}{\includegraphics{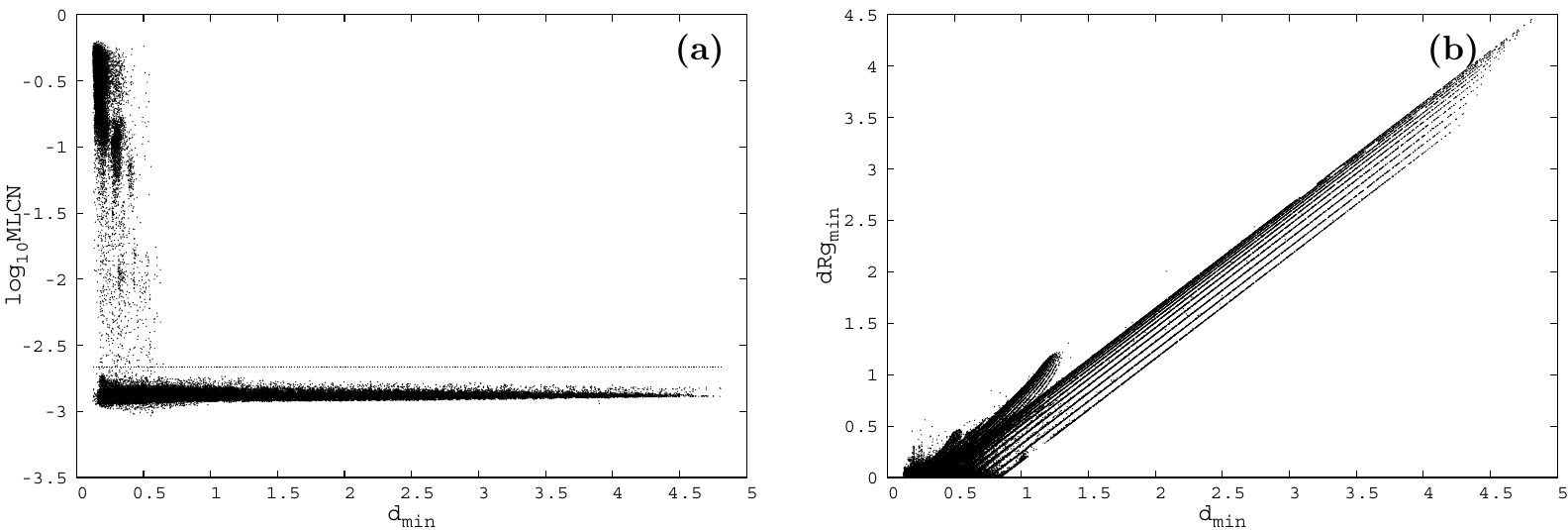}}
\caption{(a): Minimum distances of orbits to the origin versus MLCNs. The
horizontal line shows the limit separating regular from chaotic orbits.
(b): Minimum distances of orbits to the origin versus minimum
distances to the minimum of the effective potential, located at $(R_g,0)$.}
\label{dmin}
\end{figure*}

We computed, for each star of our models, their minimum distance $d_{\rm min}$
(in the meridional plane) to the origin of coordinates. Fig \ref{dmin}a
shows these minimum distances for all the orbits used to study the influence of
the mass of the bulge $M_{\rm n}$, versus the value of their respective MLCNs;
the horizontal line indicates the threshold between regular and chaotic
orbits. It is seen that \emph{all} the chaotic orbits pass near the center,
i.e., they suffer at one time or another some sudden acceleration due to the
centrifugal force. On the other hand, we can see that this is not a sufficient
condition to be chaotic: regular orbits can also pass near the center. We've
found exactly the same behaviour when using orbits from the rest of the cases
analysed in Section \ref{results}. Therefore, we may draw the following
conclusion: in the meridional plane of our galactic model, \emph{a necessary
condition for an orbit to be chaotic is to pass near the center of the
potential; a sufficient condition for an orbit to be regular is not to pass near
the center of the potential}.

From the results of the previous Section, all of the studied parameters, except
the energy, have an almost monotonic influence on the percentage of chaos in the
meridional plane. This percentage grows with the increment of the mass of the
bulge (Fig. \ref{percMn}), the decrement of the scale length of it (Fig.
\ref{perccn}), and the decrement of the angular momentum (Fig. \ref{Lzpercrit},
left). We've found (numerically) that in all the cases the position $R_g$ of the
minimum of the effective potential, which is always located on the $R$ axis
\citep[e.g.,][]{BT08}, nears the origin of coordinates whenever the percentage
of chaos rises. Fig. \ref{dmin}b shows the minimum distance to the origin
for the orbits shown in Fig \ref{dmin}a, versus their minimum distances to
$R_g$. We can see a consistent correlation between those quantities, hinting
that the position of the minimum of the effective potential might influence the
degree of chaos, although we weren't be able to find an analytic proof of this.

The picture described above is consistent with the rising of the percentage of
the chaotic motion with $M_n$ (Fig. \ref{percMn}), since the acceleration grows
with the mass of the bulge (Sec. \ref{galmod}). Also, it is consistent with the
behaviour of the chaotic percentage seen in Fig. \ref{perccn}, considering that
the more concentrated is the nucleus, the more acceleration it causes near the
center. It also explains why the percentage of chaotic motion diminishes when
the angular momentum increases (Fig. \ref{Lzpercrit}a), given that a low
angular momentum allows the star to approach the center of the potential. On the
other hand, Fig. \ref{percMn} shows that when the bulge is absent, there is no
chaotic motion at all. Whereas this proves that the onset of chaos is driven by
the presence of the nucleus, it also poses a question mark about the
abovementioned role of the centrifugal force, since it is at work even in this
full regular case.

\begin{figure*}
\resizebox{\hsize}{!}{\includegraphics{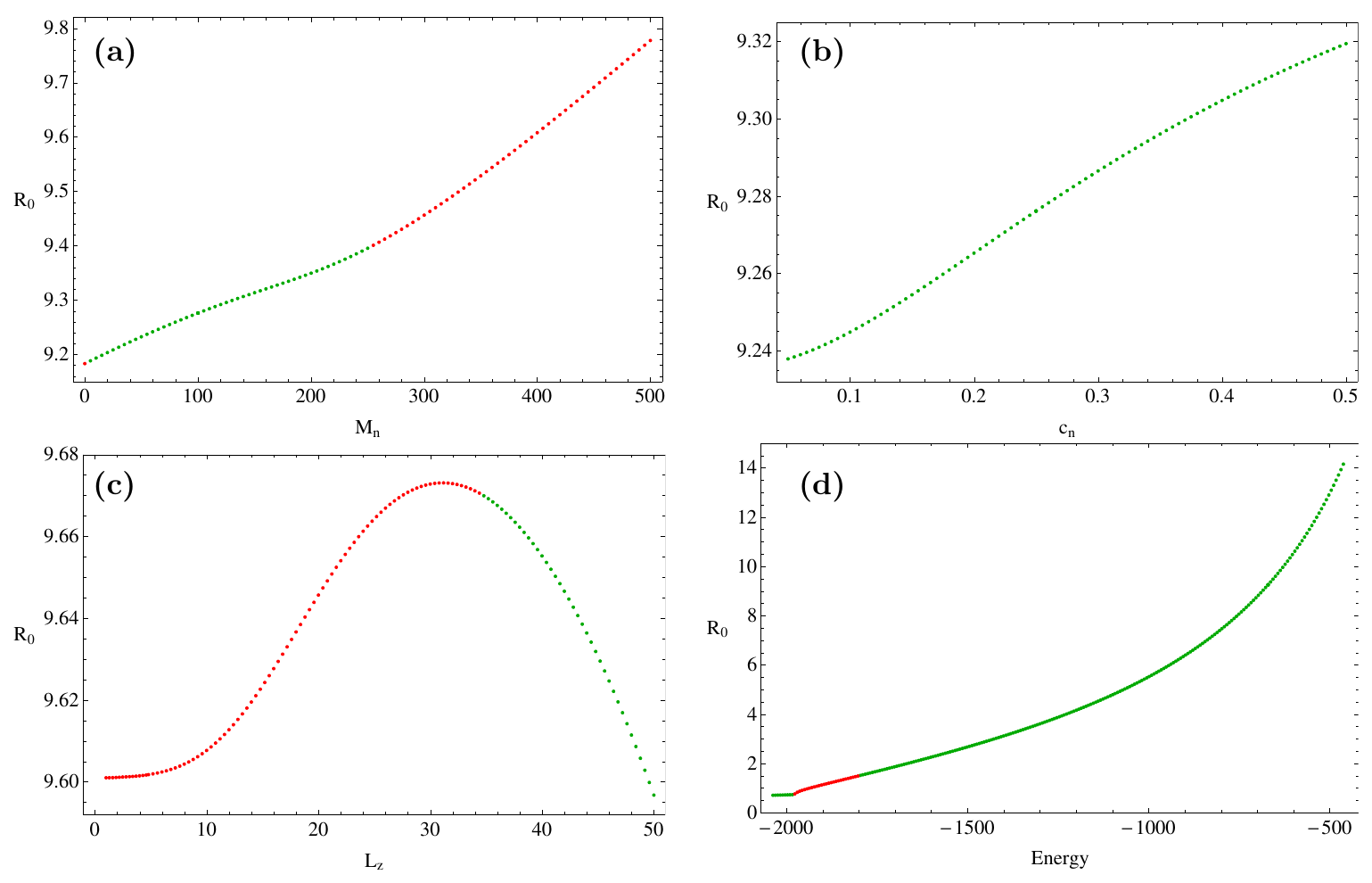}}
\caption{Evolution of the $R$ position and the stability of the 4:5
resonance varying (a) the mass of the nucleus, (b) the
scale length of the nucleus, (c) the angular momentum and (d) the energy.}
\label{Res45Evol}
\end{figure*}

On the other hand, as it was seen in the previous Section, the variation of the
parameters $M_{\rm n}$, $c_{\rm n}$, $L_z$ and $E$ causes not only variations in
the chaotic content, but also in the relative importance of the different
regular families. In particular, the 4:5 resonance may be taken as a model,
since it appears mainly when there is a low level of chaos, and decreases
significantly or even disappears when we have a strong chaotic regime.
Therefore, it is of particular interest to investigate how its stability is
being influenced by the above-mentioned parameters. For this purpose, we used the
theory of periodic orbits \citep{MH92}, in which the stability criteria can be
obtained from the elements of the monodromy matrix $X(t)$ as follows:
\begin{equation}
K = {\rm Tr} \left[X(t)\right] - 2,
\end{equation}
where Tr stands for the trace of the matrix, and $K$ is called the
\emph{stability index}. For each set of values of $M_{\rm n}$, $c_{\rm n}$,
$L_z$ and $E$, we first located, by means of an iterative process, the position
of the parent 4:5 orbit onto the $R$ axis when $\dot R=z=0$. Then, using these
initial conditions plus the value of $\dot z$ obtained from the energy, we
integrated the variational equations in order to obtain the matrix $X$, with
which we computed the index $K$. The results are presented in Figs.
\ref{Res45Evol} (a-d). In Fig. \ref{Res45Evol}a we can see the evolution of the
$R$ position and of the stability of the 4:5 resonance when $M_{\rm n}$ varies,
the values of all the other parameters being as in Fig. \ref{percMn}. Green dots
correspond to stable periodic orbits, while red dots correspond to unstable
ones. We can see that when $M_{\rm n} \geq 260$ the periodic orbit becomes
unstable. Curiously, it is also unstable when $M_{\rm n} = 0$. On the other
hand, in Fig. \ref{Res45Evol}b we see that the stability of the 4:5 parent
periodic orbit is completely unaffected by the scale length of the nucleus. In
this case, the values of all the other parameters are as in Fig. \ref{perccn}.
In Fig. \ref{Res45Evol}c, where we have used the values of the parameters as in
Fig. \ref{Lzpercrit} (left), it can been seen that there is a limit of stability
around $L_z \simeq 35$. It is worth noticing that, though in Fig. \ref{gridLz}a
there is no evidence of a 4:5 resonance, Fig. \ref{Res45Evol}c indicates that
the resonance is indeed present, although evidently deeply buried in the chaotic
sea. Finally in Fig. \ref{Res45Evol}d we present the influence of the value of
the energy. The values of all the other parameters are as in Fig. \ref{gridEnM}
(right). One may see that most of the periodic orbits are stable, except the
region $-1970 \lesssim E \lesssim -1800$ in which the periodic orbits become
unstable. It is clear, then, that the parameters of the model, as well as the
isolating integrals of motion, play a fundamental role in the stability of the
regular families, which in turn determines which ones are present in each case.

\section{Discussion and Conclusions}
\label{discus}

We have investigated how influential are the parameters of a disc galaxy with
bulge on the level of chaos and on the distribution of regular families among
its orbits. We have used an analytic axisymmetric potential which embraces the
general features of a disc galaxy with bulge, and have chosen to work in the
meridional plane of the orbits, in order to simplify the study. Varying several
of the constants of the potential, as well as the two global isolating integrals
of the orbits, namely the angular momentum and the energy, we have found that
the level of chaos and the distribution in regular families is indeed very
dependent on all of these parameters.

Our study shows that the mass of the bulge, although spherically symmetric and
therefore maintaining the axial symmetry of the whole galaxy, generates chaos in
the meridional plane as soon as it is above zero. As the mass increases, this
chaotic motion grows in percentage at the expense of the (meridional plane) box
orbits, although it seems to saturate at $\simeq 40$\% of the orbits once the
mass of the bulge has reached some $\simeq 5$\% of the mass of the disc. The
concentration of the bulge plays a similar role: the percentage of chaotic
motion depends almost linearly on this parameter. Once more, box orbits and
high-resonance orbits (by which we mean resonant orbits with a rational quotient
of frequencies made from integers $>5$) are the ones that give way to the
chaotic orbits. We also found that the angular momentum of the orbits influences
the level of chaos in a similar fashion: orbits with low angular momenta have
higher chances of being chaotic that those with high values of it. The
relationship between chaos and angular momentum is close to linear, and, again,
box orbits are the most affected by the percentage of chaos. The energy of the
orbits, however, plays a different role. In a model with low level of chaos,
varying the energy mainly shuffles the orbital content among the families of
regular orbits. Interestingly, box orbits are again the family which suffers the
most. Taking a model with a medium level of chaos, box orbits are the dominant
family at low energies, but the percentage of chaos quickly grows as the energy
increases, again by collapsing the percentage of box orbits. In this case,
however, linear orbits (i.e., 3D hollowed out saucers) grow along with chaotic
ones; also, further increasing the energy reverts these trends, and 2:1 bananas
start to increase their share. With a high level of chaos model, the increase of
the energy diminishes the percentage of chaos, while the box and 2:1 bananas
take the field.

When taken into account that the effective potential in the meridional plane has
a cusp caused by the centrifugal acceleration, all these behaviours turn out to
be consistent with the analysis we made on Sec. \ref{anres}, where we arrived to
the conclusion that a necessary condition for an orbit to be chaotic is to pass
near the center of the potential, and a sufficient condition for an orbit to be
regular is not to pass near the center of the potential.

In the same vein, we also conducted an investigation on the stability of the 4:5
resonance, which was taken as a model, in an attempt to see whether it depends
on the parameters of the galactic system and the integrals of motion. Our
results indicate that, with the exception of the scale length of the nucleus,
all the parameters affect substantially the stability of this family, hinting
at a deep interplay between chaos and proportion of regular families.

We've also found that, by combining the MLCN and the SALI algorithms, we can
obtain for both methods reliable thresholds separating chaotic from regular
motion.

We consider the outcomes of the present research as an initial effort in the
task of exploring the orbital structure of a disk galaxy with a central
spherical nucleus. Since our results are encouraging, it is in our future plans
to study the influence of all the available parameters, including the disk's
parameters $M_{\rm d}$, $\alpha$ and $h$. Moreover, we plan to obtain the entire
network of periodic orbits and reveal the evolution of their stability with
respect to all the parameters of the galactic model.

\section*{Acknowledgments}

The authors would like to thank the anonymous referee for the careful reading of
the manuscript and for all the aptly suggestions and comments which allowed us
to improve both the quality and the clarity of our work. This work was supported
with grants from the Universidad Nacional de La Plata (Argentina), the Consejo
Nacional de Investigaciones Cient\'\i ficas y T\'ecnicas de la Rep\'ublica
Argentina, and the Agencia Nacional de Promoci\'on Cient\'\i fica y
Tecnol\'ogica (Argentina).

\end{document}